\documentclass[fleqn,usenatbib]{mnras}

\usepackage{newtxtext,newtxmath}

\usepackage[T1]{fontenc}
\usepackage{ae,aecompl}

\usepackage{graphicx}	
\usepackage{amsmath}	
\usepackage{amssymb}	

\usepackage{multirow, makecell}
\usepackage{booktabs,caption}
\usepackage{threeparttable}
\usepackage{hyperref}

\title[Tracing 3D Magnetic Field Structure Using Dust Polarization]{A New Method to Trace Three-dimensional Magnetic Field Structure within Molecular Clouds Using Dust Polarization}

\author[C.-Y. Chen et al.]{
Che-Yu Chen,$^{1}$\thanks{E-mail: cc6pg@virginia.edu}
Patrick K. King,$^{1,2}$
Zhi-Yun Li,$^{1}$
Laura M. Fissel,$^{3}$
\newauthor and Renato R. Mazzei$^{1}$
\\
$^{1}$Department of Astronomy, University of Virginia, Charlottesville, VA 22904, USA\\
$^{2}$Lawrence Livermore National Laboratory, Livermore, CA 94550, USA\\
$^{3}$National Radio Astronomy Observatory, Charlottesville, VA 22904, USA
}

\date{Accepted XXX. Received YYY; in original form ZZZ}

\pubyear{2018}

\begin{document}
\label{firstpage}
\pagerange{\pageref{firstpage}--\pageref{lastpage}}
\maketitle

\begin{abstract}
The complete three-dimensional structure of the magnetic field within molecular clouds has eluded determination despite its high value in determining controlling factors in the star formation process, as it cannot be directly probed observationally.
Considering that inclination of the magnetic field relative to the plane of sky is one of the major sources of depolarization of thermal emission from dust in molecular clouds, we propose here a new method to estimate the inclination angle of the cloud-scale magnetic field based on the statistical properties of the observed polarization fraction. 
We test this method using a series of Monte Carlo experiments, and find that the method works well provided that deviations of magnetic field direction from the averaged values are small.
When applied to synthetic observations of numerical simulations of star-forming clouds, our method gives fairly accurate measurements of the mean magnetic field inclination angle (within $10^\circ-25^\circ$), which can further be improved if we restrict our technique to regions of low dispersion in polarization angles ${\cal S}$. We tested our method on the BLASTPol polarimetric observations of the Vela C molecular cloud complex, which suggests that the magnetic field of Vela C has a high inclination angle ($\sim 60^\circ$), consistent with previous analyses.

\end{abstract}




\section{Introduction}
\label{sec::intro}

Magnetic fields have long been known to play a vital role in the formation and evolution of molecular clouds (hereafter MCs), and their subsequent star formation \citep{1956MNRAS.116..503M,1991ApJ...373..169M,2007ARA&A..45..565M}.
In MCs, multi-scale supersonic flows intermittently compress material to initiate creation of a filamentary network that can be observed in both gas and dust emission \citep{2014prpl.conf...27A}. Prestellar cores arise from these turbulence-generated overdensities, and can collapse gravitationally to create protostellar systems and later become stars \citep{1987ARA&A..25...23S}. 
Magnetic effects are considered to be one of the key agents affecting the dynamics of the star forming process in MCs, in combination with turbulence and gas gravity, at all physical scales and throughout different evolutionary stages \citep{2007ARA&A..45..565M}. 
Therefore, understanding the specific roles played by magnetic fields over a range of scales is a crucial and strongly-debated topic in studies of star formation. 

It is generally recognized that non-spherical grains are oriented with their long axes perpendicular to the magnetic field lines, and thus the dust emission is linearly polarized perpendicular to the magnetic field \citep{1951ApJ...114..206D,2007JQSRT.106..225L}. 
Polarimetric observations of thermal emission from dust at mm/far-IR wavelengths are therefore used to trace the projected magnetic field orientation on the plane of sky \citep[e.g.][]{1984ApJ...284L..51H,1997ApJ...487..320N,2001ApJ...562..400M}, at least on scales larger than protostellar disks.\footnote{Other polarization mechanisms, such as scattering, may become important on the disk scale; see e.g., \cite{2015ApJ...809...78K,2016MNRAS.456.2794Y}.}

Intense observational effort has been undertaken to understand the role played by the magnetic field during star formation at different scales, including the all-sky coverage of {\it Planck} \citep[e.g.][]{2016A&A...586A.138P} and the MC-scale survey by the Balloon-borne Large Aperture Submillimeter Telescope for Polarimetry (BLASTPol; \citealt{2016ApJ...824..134F}). In addition, the Atacama Large Millimeter/submillimeter Array (ALMA), the Submillimeter Array (SMA), the James Clerk Maxwell Telescope (JCMT), and many other observatories have all successfully mapped dust polarization patterns at smaller scales in dense filaments, star-forming clumps, and protostellar envelopes and disks \citep[e.g.][]{2004ApJ...600..279C,2013ApJ...770..151C,2013ApJ...768..159H,2014Natur.514..597S,2017ApJ...838..121C,2017ApJ...842...66W}.
Despite this rich observational landscape,
it remains unclear how dynamically significant the magnetic fields are at varying physical scales, because the field strength cannot be directly measured through polarization. 
Some methods have been proposed to indirectly estimate the magnetic field strength from observations, including the widely-known Chandrasekhar-Fermi method \citep{1953ApJ...118..113C}, and the HRO (histograms of relative orientation) technique developed by \cite{2013ApJ...774..128S}, which uses the shape of the distribution of relative orientations between the gas structure and polarization vectors to determine the relative importance of the magnetic field. 
Following this direction,
\citet[][hereafter \hyperlink{CKL16}{CKL16}]{2016ApJ...829...84C} further proposed that the transition of the shape of HROs based on gas densities could be used to infer the total magnetic field strength of the cloud.

More fundamentally, the 3D structure of magnetic field (e.g.~ratio between line-of-sight and plane-of-sky field components) is still unknown in nearby star-forming regions.
By comparing with synthetic observations of numerical simulations, the statistical power of polarization data can in principle become an important factor to constrain the direction of the MC-scale magnetic field.
For example,
\citet[][hereafter \hyperlink{KFCL18}{KFCL18}]{2018MNRAS.474.5122K} compared their star-forming, MC-scale simulations with the BLASTPol polarimetric data from the Vela C cloud \citep{2016ApJ...824..134F} through detailed statistical studies, and concluded that the magnetic field direction within Vela C might be very close to our line of sight.

In this manuscript, we introduce a new yet simple method to estimate the overall inclination angle of the MC-scale magnetic field, based on the measured polarization fraction of thermal dust emission from the cloud. 
The intrinsic polarization coefficient $p_0$ (assumed to be constant across the cloud) can be derived from the measured maximum polarization fraction $p_\mathrm{max}$ (which must be generated by a line of sight with nearly uniform magnetic fields almost completely on the plane of sky) within the cloud.
Under the assumption of perfect grain alignment,
when the dispersion of the plane-of-sky magnetic field direction is small, the inclination angle becomes the dominant source of depolarization. 
If cancellation within the line-of-sight can be neglected, one may obtain an estimate for the inclination angle of the magnetic field at every pixel from the observed polarization map. 
The characteristic inclination angle of the whole cloud can then be determined statistically using the location of the peak value of the probability distribution function (PDF) of these estimates, which is calculated using Gaussian kernel density estimation (KDE).

This paper is organized as follows. In Section~\ref{sec::pol} we briefly review the basic equations for calculating the Stokes parameters of polarized emission from dust (Section~\ref{sec::poleqs}), and show the derivation, from these equations, of the $p$-estimated magnetic field inclination angle $\gamma_\mathrm{obs}$ (Section~\ref{sec::derive}). A Monte Carlo study is presented in Section~\ref{sec::MC} as an independent test on the analytical method. We applied our method to fully-3D cloud-scale simulations in Section~\ref{sec::test}, where we performed two sets of tests on varying cloud environment (Section~\ref{sec::clouds}) and viewing angles (Section~\ref{sec::rot}). Finally, in Section~\ref{sec::velac} we applied this new method on the Vela C molecular cloud using the polarization data from BLASTPol. We summarize our work in Section~\ref{sec::sum}.

\section{Dust Polarization}
\label{sec::pol}

\subsection{Basic Equations}
\label{sec::poleqs} 

\begin{figure}
\centering
	\includegraphics[width=0.8\columnwidth]{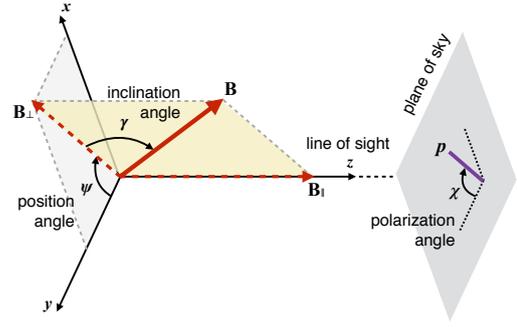}
    \caption{Definition of symbols. This is an updated version of Figure~1 of \protect\hyperlink{CKL16}{CKL16}.}
    \label{angdef}
\end{figure}

Here we review the widely adopted dust polarization equations based on previous work \citep[e.g.][]{1985ApJ...290..211L,2000ApJ...544..830F,2013ApJ...774..128S,2015A&A...576A.105P}. Symbols are defined in Figure~\ref{angdef}, which is adapted from \hyperlink{CKL16}{CKL16}.
The synthetic polarization is determined from the integrated Stokes parameters using the volume density $n$ and magnetic field $B \equiv |\mathbf{B}| = |B_x \hat{\mathbf{x}} + B_y \hat{\mathbf{y}} + B_z \hat{\mathbf{z}} |$:
\begin{align}
q &= \int n \frac{B_y^2 - B_x^2}{B^2} ~dz = \int n \cos 2\psi \cos^2 \gamma ~dz , \notag\\
u & = \int n \frac{2B_xB_y}{B^2} ~dz = \int n \sin 2\psi \cos^2 \gamma ~dz,
\label{eq::qu}
\end{align}
where $\gamma$ is the inclination angle with respect to the plane of sky and $\psi$ is the position angle on the plane of sky (see Figure~\ref{angdef}). The polarization fraction is
\begin{equation}
p = p_0 \frac{\sqrt{q^2 + u^2}}{N - p_0 N_2},
\end{equation}
where $N = \int n ~dz$ is the column density integrated along the line of sight, and
\begin{equation}
N_2 = \int n\left(\cos^2 \gamma - \frac{2}{3}\right) ~dz
\label{eq::polN2}
\end{equation}
is a correction term considering reduced emission from inclined dust grains with smaller cross-section \citep{2000ApJ...544..830F}. 
$p_0$ is a coefficient determined by dust grain properties, and is assumed to be constant throughout a cloud.
The inferred polarization angle on the plane of sky is given using the four-quadrant arctangent
\begin{equation}
\chi = \frac{1}{2} \text{arctan2}(u,q).
\end{equation}

The dispersion in polarization angles, ${\cal S}$, is usually defined at each pixel as the averaged difference between the direction of polarization vectors 
at this pixel $x$ and other pixels $x_i$ located a distance $\delta$:
(see e.g.~\citealt{2008ApJ...679..537F, 2015A&A...576A.104P,2016ApJ...824..134F}; \hyperlink{KFCL18}{KFCL18})
\begin{equation}
{\cal S}^2 (x, \delta) = \frac{\sum \Delta\chi^2(x, x_i)}{\mathrm{number\ of} x_i}, \label{eq::polS}
\end{equation}
where the angular difference in polarization between pixels $x$ and $x_i$ can be written as
\begin{align}
\Delta\chi(x, x_i) = \frac{1}{2}\text{arctan2}\big( & q(x_i) u(x) - q(x) u(x_i),\notag \\
& q(x_i) q(x) + u(x_i) u(x) \big),
\end{align}
which can be directly calculated from the Stokes parameters. 
The inclusion of $\delta$ (and therefore the form of correlation function for ${\cal S}$) is more relevant in observational data, where the spatial resolution/telescope beam size must be taken into consideration when calculating the dispersion.
In this study, we only consider the dispersion measured among the 8 nearest neighbors of a given pixel (see \hyperlink{CKL16}{CKL16}) to utilize the intrinsically high resolution of numerical simulations, but we caution the readers that dispersion could be a function of scale (see also Section~\ref{sec::velac}).

\subsection{Inclination Angle and Polarization Fraction}
\label{sec::derive}

As discussed in \hyperlink{CKL16}{CKL16}, the polarization fraction is determined by two major factors: inclination angle $\gamma$, and the dispersion of position angle $\psi$ along the line of sight, because this dispersion gives rise to cancellation within the line-of-sight. 
If we consider the ``perfect'' scenario when there is no variation in neither $\gamma$ or $\psi$ along the line of sight, the polarization fraction from Equations~(\ref{eq::qu})$-$(\ref{eq::polN2}) becomes
\begin{equation}
p = \frac{p_0\cos^2\gamma}{1-p_0\left(\cos^2\gamma-\frac{2}{3}\right)}.
\label{eq::pobs}
\end{equation}
Theoretically, the maximum value of $p$ happens when $\cos^2\gamma =1$ ($\gamma = 0$; the magnetic field is completely on the plane of sky). 
The maximum polarization fraction one can measure in a cloud is therefore \begin{equation}
p_\mathrm{max} = \frac{p_0}{1 - \frac{1}{3}p_0},
\end{equation}
or equivalently,
\begin{equation}
p_0 = \frac{3 ~p_\mathrm{max}}{3+p_\mathrm{max}}.
\label{eq::p0}
\end{equation}
This provides a way to estimate the polarization coefficient $p_0$,
which is directly related to dust grain properties of the cloud \cite[size distribution, alignment efficiency, etc.; see e.g.][]{2007JQSRT.106..225L}. Note that, as we discussed in Appendix~\ref{sec::appx}, 
$p_{\rm max}$ can be recovered quite accurately in general (although some exceptions exist), 
and therefore Equation~(\ref{eq::p0}) can be used to determine $p_0$ within gas structures of different scales (clouds, clumps, dense cores, etc.). Such information can potentially provide a powerful probe of the dust grain properties among various physical scales and at different evolutionary stages during star formation. 

Once we have $p_0$ derived from the measured $p_\mathrm{max}$, we can calculate $\cos^2\gamma_\mathrm{obs}$ explicitly from the observed polarization fraction $p_\mathrm{obs}$ using Equation~(\ref{eq::pobs}):
\begin{equation}
\cos^2\gamma_\mathrm{obs} =\frac{ p_\mathrm{obs}\left(1+\frac{2}{3}p_0\right)} {p_0\left(1+p_\mathrm{obs}\right)}.
\label{eq::cosrobs}
\end{equation}
As a result, $\gamma_\mathrm{obs}$ is the inclination angle from the plane of sky corresponding to $p_\mathrm{obs}$, and can be derived at every pixel from any polarimetric observation map. 

However, note that we assumed no cancellation along the line-of-sight from variations of magnetic field directions, and that the inclination is the only source of depolarization (i.e.~homogeneous grain alignment). The angle $\gamma_\mathrm{obs}$ derived from Equation~(\ref{eq::cosrobs}) therefore also represents the largest-possible inclination angle corresponding to $p_\mathrm{obs}$ and $p_0$. Because of these uncertainties, when referring to the cloud-scale magnetic field orientation, we consider the most probable value of $\gamma_\mathrm{obs}$ of all detections of $p_\mathrm{obs}$ among the entire cloud, denoted as ${\gamma_\mathrm{obs}}^\wedge$, to potentially reduce the errors statistically.
Since the clouds are expected to mostly be background with a few regions containing overdense structures like filaments and cores, the peak location of PDF, or the most probable value, could in principle represent this background value, which is what we are looking for: the inclination angle of cloud-scale magnetic field.
Nevertheless, we would like to point out that in the case of a highly disordered magnetic field, it will be intrinsically difficult to characterize the magnetic field structure by just one direction, and therefore the error in the derived inclination angle from Equation~(\ref{eq::cosrobs}) increases with the level of perturbation within the cloud (see Appendix~\ref{sec::appx} for more details). 
Also note that the most probable value is only significant with a large number of measurements to provide sufficient statistical coverage, which are exactly the cases of numerical simulations and the BLASTPol data of Vela C (see Section~\ref{sec::velac}) considered in this study.
We perform various tests to investigate the accuracy of this $p$-derived inclination angle in the following sections.

\section{Monte Carlo Experiments}
\label{sec::MC}

\begin{figure}
\includegraphics[width=\columnwidth]{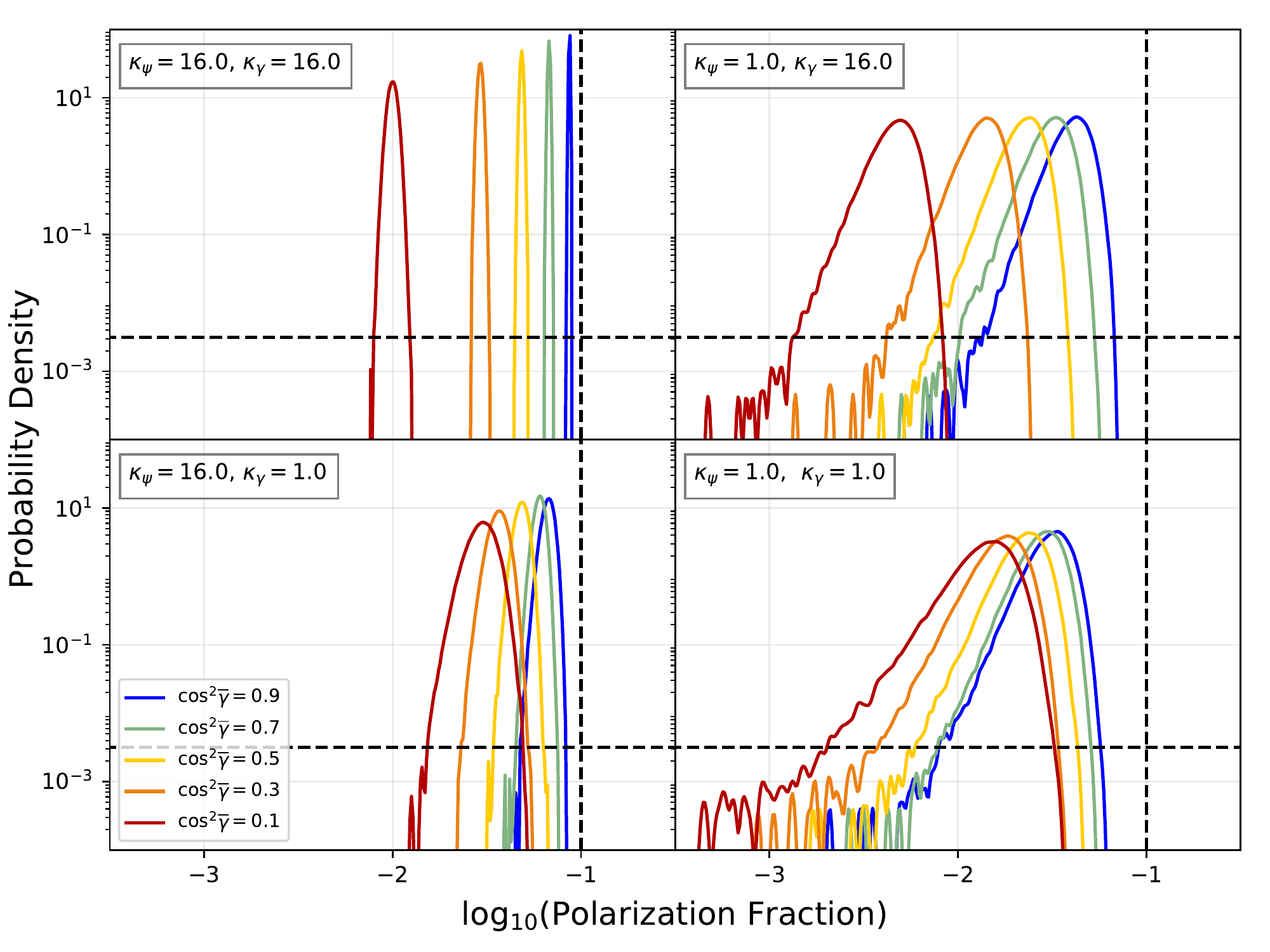}
\vspace{-.2in}
\caption{Polarization fraction PDFs obtained using our Monte Carlo 
model. Each panel contains five Monte Carlo simulations, each with a 
different choice for $\cos^2 \overline{\gamma}$. The top row contains 
simulations with $\kappa_{\gamma} = 16$ (highly concentrated $\gamma$ values), and the bottom row, simulations with $\kappa_{\gamma} = 1$ (highly disordered $\gamma$ values); the left column contains simulations with $\kappa_{\psi} = 16$, and the right column, simulations with $\kappa_{\psi} = 1$. Vertical dashed lines represent the polarization coefficient adopted in these Monte Carlo experiments, $p_0 = 0.1$, and horizontal dashed lines mark the $N^{-1/2}$ error in Monte Carlo simulations.
The peaks of the PDFs change accordingly with $\cos^2\overline{\gamma}$, though the locations of the peaks (see Table~\ref{MCcomp}) as well as the shapes of the distributions could be significantly altered because of the highly disturbed values of either $\gamma$ or $\psi$.}
\label{MCplot}
\end{figure}

To investigate the accuracy of our analytical correlation of $p$ and $\gamma$, we conducted Monte Carlo experiments to study the effects of parametrized statistical perturbations on the theoretical estimate, Equation~(\ref{eq::pobs}). 
The main purpose of this set of tests is to examine whether, in less-complicated systems and with better-quantified distortions, the dependence of the peak location of $p$ distribution on inclination angle $\gamma$ agrees with the analytic prediction described in Section~\ref{sec::derive}. 

Our Monte Carlo model for the polarization fraction consists of a set of $N$ realizations of polarized emission due to a set of $M$ random unit vectors (corresponding to magnetic field orientation within a single line-of-sight) which are used to generate a pair of Stokes parameters from Equation~(\ref{eq::qu}). This generates an ensemble of $N$ polarization fractions, upon which KDE is used to estimate the polarization fraction PDF (see e.g.~\hyperlink{KFCL18}{KFCL18}). For simplicity, we assume in these Monte Carlo experiments that the gas density is the same everywhere. 

To sample our random vectors, we assume that
they are described by two independent distributions for the inclination angle $\gamma$ and the plane-of-sky position angle $\psi$ (see Figure~\ref{angdef}). 
The von Mises distribution is a well-known analogue of the Gaussian for circular variables \citep{1995sacd.book.....F}, and is a suitable parametric choice for these orientation angles for random vectors centered at an average quantity subject to statistical perturbation. For some angle $\theta$ with circular mean $\overline{\theta}$ and concentration $\kappa$, the von Mises distribution for $\theta$ is
\begin{equation}
P(\theta)\big|_{(\overline{\theta} , \kappa)} = \frac{e^{\kappa\cos(\theta - \overline{\theta})}}{2\pi\cdot I_0(\kappa)}
\end{equation}
where $I_0$ is the zeroth order modified Bessel function. The concentration, $\kappa$, goes roughly as the inverse of the variance ($\kappa \sim \sigma^{-2}$ where $\sigma$ is the standard deviation). Since the observer is free to rotate their coordinate system on the plane-of-sky such that $\overline{\psi} = 0$, we are free to simplify our model by observing that for any choice of $\overline{\psi}$, provided that the sightline in question is taken out of context with its neighbors, we can rotate our sightline such that $\overline{\psi}=0$. Under this assumption, each Monte Carlo simulation depends only on the three parameters, $\overline{\gamma}$ (the mean inclination angle), $\kappa_{\gamma}$ (the concentration of the inclination angle), and $\kappa_{\psi}$ (the concentration of the position angle).

We conducted Monte Carlo simulations for five mean inclination angles ($\cos^2 \overline{\gamma} = 0.9, 0.7, 0.5, 0.3, 0.1$), two inclination angle concentrations ($\kappa_{\gamma} = 1, 16)$, and two position angle concentrations ($\kappa_{\psi} = 1, 16$). For each simulation, we adopted $M = 50$ (the number of random vectors per sightline) and $N = 10^5$ (number of polarization fraction samples) to be comparable with synthetic observations discussed in Section~\ref{sec::test}. The resulting KDE estimates of the polarization fraction PDFs are presented in Figure~\ref{MCplot}. The error in Monte Carlo simulations goes as $N^{-1/2}$, which is annotated as a dashed black horizontal line in each plot; the adopted polarization fraction coefficient $p_0 = 0.1$ is also annotated as a vertical line.

\begin{figure}
\centering
\includegraphics[width=\columnwidth]{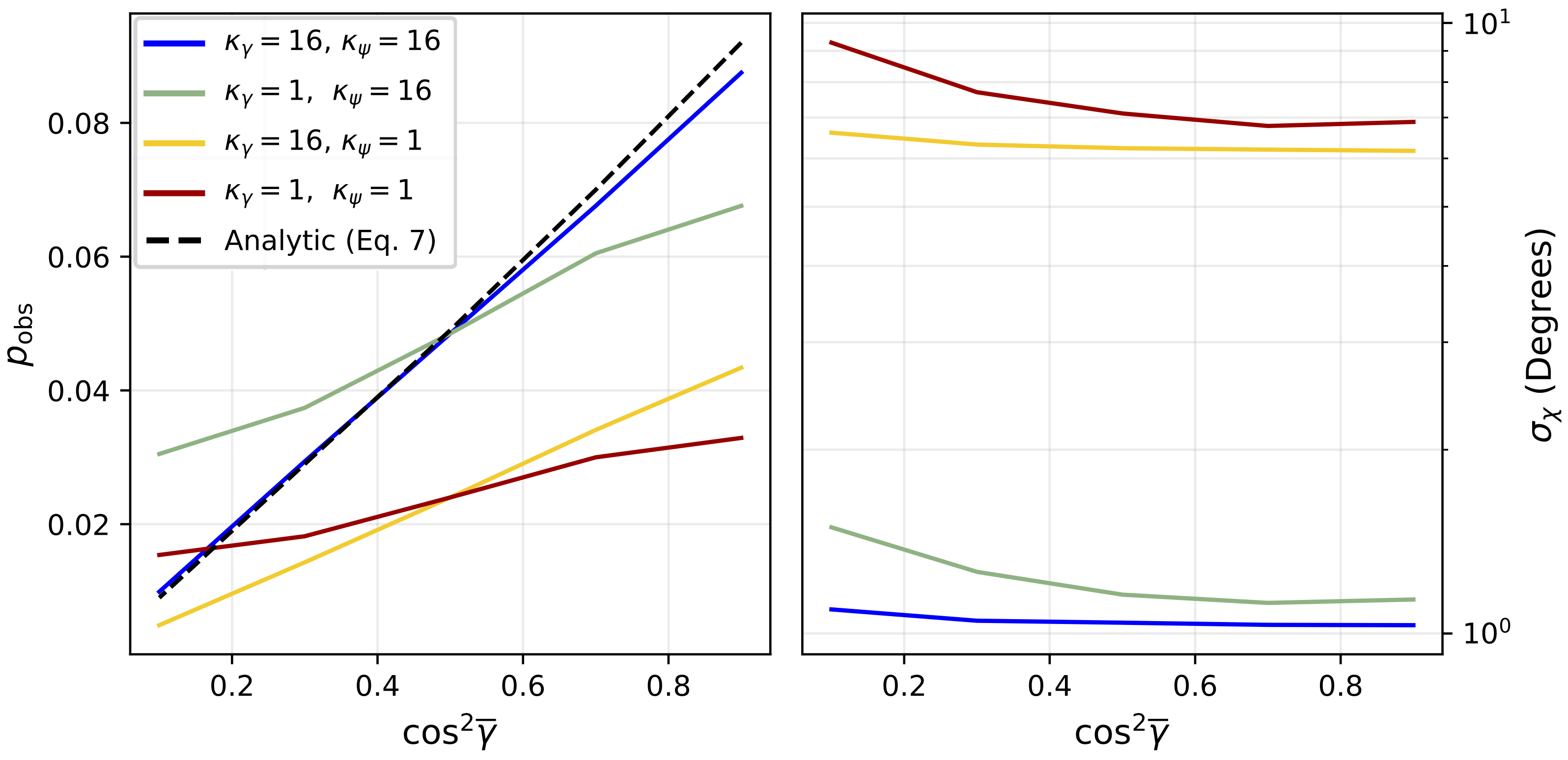}
\vspace{-.2in}
\caption{{\it Left:} Plot of $p_\mathrm{obs}$ as functions of $\cos^2 \overline{\gamma}$, for both the analytic solution (Equation~(\ref{eq::pobs}); {\it blask dashed line}) and the peak locations measured from our Monte Carlo models in Figure~\ref{MCplot}.
{\it Right:} the circular standard deviation of the polarization angle, $\sigma_\chi$, measured in out Monte Carlo models with different mean inclination angles.
Note the excellent agreement between the analytic prediction for the $\kappa_{\gamma} = 16$, $\kappa_{\psi} = 16$ simulation ({\it blue line}), which also has the lowest $\sigma_{\chi}$. Decreasing 
$\kappa_{\gamma}$ worsens agreement with the analytic prediction, especially at extreme values of $\cos^2 \overline{\gamma}$ (magnetic field being roughly parallel or perpendicular to the line of sight). 
Models with wide distribution of position angle $\psi$ (concentration level $\kappa_\psi=1$; {\it yellow} and {\it red} lines) also have large uncertainties in polarization angle ($\sigma_\chi\sim 6^\circ-10^\circ$), which result in significant deviation from the theoretical prediction of the $p_\mathrm{obs} - \cos^2\overline{\gamma}$ correlation. }
\label{fig::MCcomp}
\end{figure}

We used the peak of the polarization fraction PDF (i.e.~the most probable value) as an estimator for $p_\mathrm{obs}$ for each simulation, which is presented in the left panel of Figure~\ref{fig::MCcomp} as a function of $\cos^2 \overline{\gamma}$. It is immediately apparent that the Monte Carlo simulations with the smallest amplitude perturbations to either $\gamma$ or $\psi$ result in the best agreement with the analytic estimates (the dashed black line). Reducing the concentration $\kappa_{\gamma}$ to 1 induces the Monte Carlo simulations to deviate more strongly from the analytic estimates, particularly as $\cos^2 \overline{\gamma}$ varies far from the central value of 0.5. 
Reducing $\kappa_{\gamma}$ appears to have the same effect as the extreme inclination effects at low $\kappa_{\psi}$. 
These results suggest that as the concentration $\kappa_{\gamma}$ weakens, the value of $\cos^2 \overline{\gamma}$ as an overall measure of inclination angle is reduced considerably. Extreme inclination values are therefore far more rarely realized, and as a result, the overall measured inclination angle should be driven away from such extreme values, even though the distributions themselves are centered at those values. This is consistent with the interpretation in \hyperlink{KFCL18}{KFCL18} of inclination as a mixing angle, which is not necessarily representative of a coherent inclination shared by all sightlines in any particular target.

Reducing the concentration of $\kappa_{\psi}$ drives the Monte Carlo simulation results far from the simple analytic predictions, particularly at high $\cos^2 \overline{\gamma}$ (small inclination angle; magnetic field almost parallel to the plane of sky), when reductions in polarization fraction due to inclination effects are expected to be minimized (also see discussion in Section~\ref{sec::rot}). This suggests that at low concentration $\kappa_{\psi}$ the effects of cancellation within the line-of-sight, where polarization signals interfere destructively, becomes more important. One indirect probe of cancellation effects is the dispersion in polarization angles $\mathcal{S}$ (see Equation~(\ref{eq::polS})): as a population, measurements of $\mathcal{S}$ in regions with high cancellation should be higher on average than regions with low cancellation, neglecting coherent structures in $\mathcal{S}$ that arise due to dominant magnetohydrodynamical flows. Our Monte Carlo simulations are manifestly unable to compute $\mathcal{S}$ as each sightline is considered independently of other sightlines; however, a polarization angle ($\chi$; see Figure~\ref{angdef}) can be computed from the Stokes parameters used to compute the polarization fraction, and these may be considered as a population. We use the circular standard deviation of the polarization angles, $\sigma_\chi$, as an estimator for $\mathcal{S}$. These are presented in the right panel of Figure~\ref{fig::MCcomp}. Those Monte Carlo simulations with high concentration $\kappa_{\psi}$ tend to have the lowest $\sigma_\chi$, and thus might be expected to correspond to regions with the lowest $\mathcal{S}$. 
We therefore conclude that the analytical solution, Equation~(\ref{eq::cosrobs}), works better when the dispersion in magnetic field direction on the plane of sky is small, and when the magnetic field is neither completely on the plane of sky or perfectly along the line of sight. This leads to our analysis in Section~\ref{sec::test} below.

\begin{table}
  \begin{threeparttable}
	\caption{Comparison between the analytical result and Monte Carlo test, both with $p_0 = 0.1$.}
	\label{MCcomp}
	\begin{tabular}{c | c | cccc}
	\hline
	 \multirow{4}{*}{$\cos^2\gamma\ (\gamma)$} & \multicolumn{5}{c}{estimated/measured polarization fraction $p$}\\
	 \cline{2-6}
	 & Analytical & \multicolumn{4}{c}{Monte Carlo Method$^{\dagger}$}\\
	\cline{3-6}
	  & Solution & $\kappa_\gamma = 16$, & $\kappa_\gamma = 1$, & $\kappa_\gamma = 16$, & $\kappa_\gamma = 1$, \\
	  & (Eq.~(\ref{eq::pobs})) & $\kappa_\psi = 16$ & $\kappa_\psi = 16$ & $\kappa_\psi = 1$ & $\kappa_\psi = 1$ \\
	\hline
	        0.9 ($18.4^\circ$) & 0.092 & 0.088 & 0.068 & 0.043 & 0.038  \\
		0.7 ($33.2^\circ$) & 0.070 & 0.068 & 0.061 & 0.034 & 0.030 \\
		0.5 ($45.0^\circ$)  & 0.049 & 0.049 & 0.049 & 0.024 & 0.024 \\
		0.3 ($56.8^\circ$)  & 0.029 & 0.029 & 0.037 & 0.014 & 0.018 \\
		0.1 ($71.6^\circ$)  & 0.009 & 0.010 & 0.031 & 0.005 & 0.015 \\
		\hline
	\end{tabular}
    \begin{tablenotes}
      \footnotesize
      \item $^\dagger$Columns (3)$-$(6) are values measured at the peaks of the PDFs, ${p}^\wedge$.
    \end{tablenotes}
  \end{threeparttable}
\end{table}

\section{Numerical Validation}
\label{sec::test}

\begin{figure*}
\includegraphics[width=\textwidth]{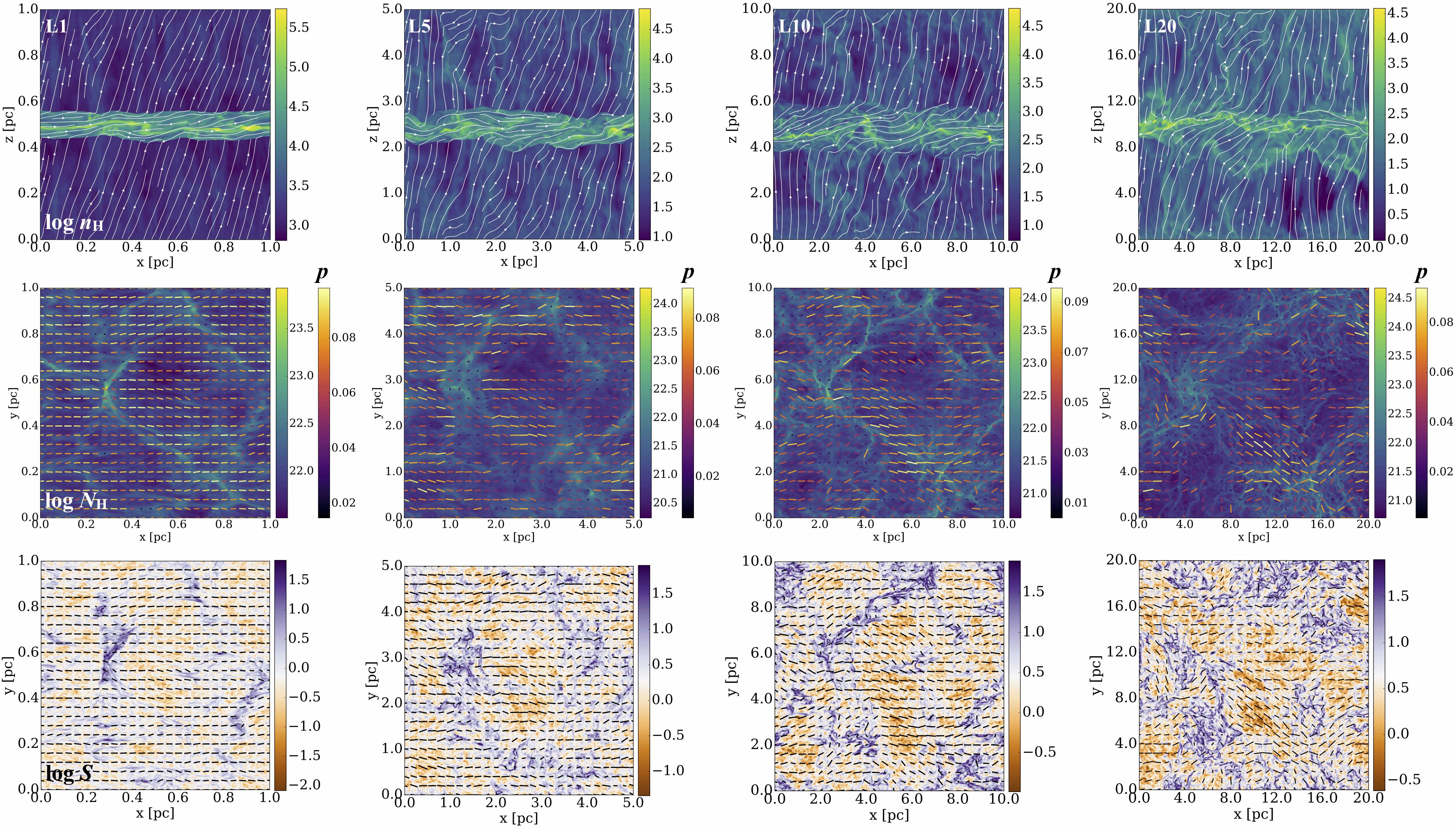}
\vspace{-.2in}
\caption{The structure of simulated clouds. {\it Top row:} number density of hydrogen nucleus $n_\mathrm{H}$ (colormap, in $\log$-scale) and magnetic field structures (white streamlines) of a slice cut through $y\approx 0.4~L_\mathrm{box}$. Note that the pre-shock regions are not included in the synthetic observations and following analysis. {\it Middle row:} column density $N_\mathrm{H}$ (colormap, in $\log$-scale) integrated through $z$-axis with synthetic polarization vectors (color-coded by polarization fraction). {\it Bottom row:} maps of dispersion of polarization angle ${\cal S}$ with synthetic polarization vectors.  The colorscales of the ${\cal S}$ maps are centered at the median value, $\langle{\cal S}\rangle$; in principle, regions with ${\cal S} < \langle{\cal S}\rangle$ (i.e.~brown-yellow regions) could give better estimate of the magnetic field inclination angle $\gamma$ (see Table~\ref{simprop}). }
\label{fig::cloudsim}
\end{figure*}

\begin{table*}
	\centering
  \begin{threeparttable}
	\caption{Major properties of simulations considered in this study, as well as the polarimetric properties and polarization-inferred inclination angles from synthetic observation of these models along the $z$ direction. The superscript $^\wedge$ represent the most probable values of that property (i.e.~the peak locations of PDFs), and $\sigma$ represents the standard deviation. Results from Vela C cloud are also included for comparison.}
	\label{simprop}
	\begin{tabular}{l | cccc | cccc | cc | ccccc}
		\hline
		\multirow{2}{*}{model} & $L_\mathrm{box}$ & $v_\mathrm{rms}$$^{\dagger}$ & \multirow{2}{*}{$\beta_\mathrm{plasma}$$^{\dagger}$} & \multirow{2}{*}{${\cal M}_\mathrm{A}$$^{\dagger}$} & \multirow{2}{*}{${p}^\wedge$} & \multirow{2}{*}{$\sigma_{\log p}$} & \multirow{2}{*}{$\langle{\cal S}\rangle$} & \multirow{2}{*}{$\sigma_{\log {\cal S}}$} & \multirow{2}{*}{$p_\mathrm{max}$} & derived & $\gamma_{\overline{\mathbf{B}}}$ & ${\gamma_\mathrm{3D}}^\wedge$ & ${\gamma_\mathrm{2D}}^\wedge$ & ${\gamma_\mathrm{obs}}^\wedge$ & ${\gamma_{\mathrm{obs}_,\ {\cal S}<\langle{\cal S}\rangle}}^\wedge$ \\
& (pc) & (km~s$^{-1}$) & & & & & & & & $p_0$ & $(^\circ)$ & $(^\circ)$ & $(^\circ)$ & $(^\circ)$ & $(^\circ)$ \\
\hline
L1 & $\ \ 1$ & 0.76 & 0.15 & 1.04 & 0.089 & 0.09 & 0.8$^\circ$ & 0.36 & 0.101 & 0.098 & \ \ 7.7 & $1.4$ & 11.2 & 20.6 & 18.5\\
L5 & $\ \ 5$ & 0.99 & 0.14 & 1.33 & 0.068 & 0.18 & 1.9$^\circ$ & 0.37 & 0.096 & 0.093 & \ \ 8.7 & 2.2 & 17.5 & 39.0 & 31.9 \\
L10 & 10 & 1.87 & 0.05 & 1.45 & 0.067 & 0.22 & 2.8$^\circ$ & 0.41 & 0.100 & 0.097 & 17.6 & 9.7 & 19.8 & 38.8 & 34.8  \\
L20 & 20 & 2.15 & 0.03 & 1.39 & 0.060 & 0.25 & 4.4$^\circ$ & 0.41 & 0.100 & 0.097 & 15.2 & 4.0 & 19.6 & 44.0 & 39.3 \\
\hline
Vela C & $-$ & $-$ & $-$ & $-$ & 0.043 & 0.28 & 7.9$^\circ$ & 0.28 & 0.150 & 0.142 & $-$ & $-$ & $-$ & 64.9 & 54.5 \\
\hline
	\end{tabular}
    \begin{tablenotes}
      \footnotesize
      \item $^\dagger$We note that these measurements only serve as references, not definite properties of individual clouds. Since every MC is spatially large and could cover a wide range of physical environments, it is inappropriate to use a single value to represent the entire cloud. Also, note that though the Alfv{\'e}n Mach number ${\cal M}_\mathrm{A}$ gives the ratio between kinetic and magnetic energies, it does not capture the fact that most of the post-shock flows are along the magnetic field lines (and therefore do not lead to distortion of magnetic field structure).
    \end{tablenotes}
  \end{threeparttable}
\end{table*}

\subsection{Simulations}
\label{sec::defgamma}

We next tested our method
on synthetic observations of simulated star-forming molecular clouds.
Using {\it Athena} \citep{2008ApJS..178..137S}, we conducted 4 fully-3D MHD simulations to cover a range of physical properties in the simulated clouds (see Table~\ref{simprop}), which are shown in Figure~\ref{fig::cloudsim}. 
The simulation setup considers the commonly adopted cloud-cloud collision scenario of MC formation \citep[e.g.][]{2006ApJ...648.1052H, 2006ApJ...643..245V, 2013ApJ...774L..31I, 2014prpl.conf....3D,2017ApJ...847...140C}; 
we constructed dense, star-forming regions via supersonic collision of two diffuse, turbulent, and magnetized clouds. 
In our simplified scenario, we assumed our simulation box is just big enough to cover the colliding clouds ($L_\mathrm{box} \sim 2 R_\mathrm{cloud}$); therefore, plane-parallel convergent flows (along $z$-direction) from both sides of the box are added on top of the local turbulence as the inter-cloud velocity between the two clouds to induce the collision. We also set the initial magnetic fields in both clouds to be at $20^\circ$ with respect to the convergent flows to create oblique MHD shocks; for simplicity, we assumed both clouds have the same magnetic field strength. 
As described in \citealt{2012ApJ...744..124C,2014ApJ...785...69C,2015ApJ...810..126C} (hereafter \hyperlink{CO12}{CO12}, \hyperlink{CO14}{CO14}, \hyperlink{CO15}{CO15})
and \cite{2017ApJ...847...140C}, these flows compress gas to form a post-shock layer, which has density and magnetic field strength comparable to those observed in MCs. 
Following the general equations of MHD shocks, the desired MC conditions (magnetically supercritical, super-Alfv{\'e}nic, etc.) can be easily achieved by selecting appropriate inflow conditions (see derivations in  \hyperlink{CO12}{CO12} and  \hyperlink{CO14}{CO14}).

Since the bulk motion in these convergent flows is neutralized at the shock front, local turbulence within the two colliding clouds becomes important kinematically in the post-shock region. Following \hyperlink{CO14}{CO14} and \hyperlink{CO15}{CO15}, we generate the velocity perturbation that follows a power spectrum ${v_k}^2 \propto k^{-4}$ based on the observational results from MCs (\citealt{2007ARA&A..45..565M}; also see e.g.~\citealt{2011ApJ...729..120G,2015ApJ...806...31G}). The velocity perturbation amplitudes within the colliding clouds are determined by assuming that these clouds are viralized with virial number $\alpha_\mathrm{vir}=2$:
\begin{equation}
\alpha_\mathrm{vir} \equiv \frac{5{\sigma_v}^2 R_\mathrm{cloud}}{G M_\mathrm{cloud}} = 2.
\end{equation}
With $M_\mathrm{cloud} \sim 4\pi \rho_0 {R_\mathrm{cloud}}^3/3$ and $L_\mathrm{box} \sim 2 R_\mathrm{cloud}$, the local velocity amplitude from turbulence is
\begin{equation}
\sigma_v = \sqrt{\frac{G\pi\alpha_\mathrm{vir}}{15}}\cdot {\rho_0}^{1/2} \cdot L_\mathrm{box}.
\end{equation}
We chose $\alpha_\mathrm{vir} = 2$ and $\rho_0 = 50~\mathrm{cm}^{-3}$ for all of our cloud-scale simulations. 

Table~\ref{simprop} lists the major properties of the dense clouds formed in our simulations,\footnote{We only listed the post-shock conditions of these simulations and not the pre-shock parameters, because the key purpose of these simulations is to create star-forming environments similar to those observed in MCs, not to investigate the formation mechanism of MCs. The initial conditions of our simulations are therefore less relevant to this study. } measured at the time when the gas density reaches $n_\mathrm{max} \geq 10^7$~cm$^{-3}$ (a condition considered as the formation of protostars; see \hyperlink{CO14}{CO14} and \hyperlink{CO15}{CO15} for more discussions).
Roughly speaking, models L1 to L20 follow the trend from more magnetically-dominated to more turbulent (see Figure~\ref{fig::cloudsim}).
Note that model L10 is previously reported as model A in \hyperlink{KFCL18}{KFCL18}.
Also, model L1 is actually a smaller-scale, prestellar core-forming simulation from \hyperlink{CO15}{CO15} (their model M10B10); for this type of ``local" box embedded in MCs, the convergent flows are representing the cloud-scale ($\gg L_\mathrm{box}$) turbulence, and the velocity perturbation amplitude follows the scaling relation $\delta v \propto \ell^{1/2}$ (see Equation~(21) of \hyperlink{CO14}{CO14}).\footnote{In this setup, there is separation of scales between the box size and the cloud scale corresponding to the convergent flow speed according to the Larson's law, and turbulence information in-between is missing in the simulation. }

\begin{figure}
\centering
\includegraphics[width=\columnwidth]{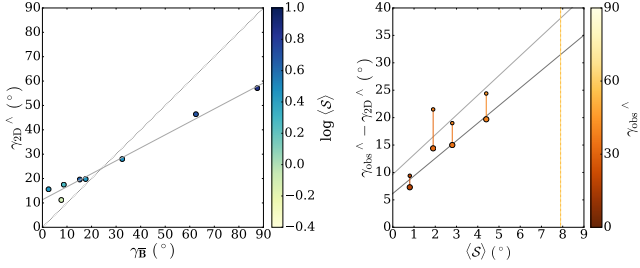}
\vspace{-.2in}
\caption{{\it Left:} Comparison of the most probable values of 2D projected inclination angle ${\gamma_\mathrm{2D}}^\wedge$ with the inclination angle of the average magnetic field in the 3D space $\gamma_{\overline{\mathbf{B}}}$, for all tests discussed in Sections~\ref{sec::clouds} and \ref{sec::rot}. All dots are color-coded by the polarization angle dispersion (in $\log$ space) of corresponding models; the perfect scenario ${\gamma_\mathrm{2D}}^\wedge = \gamma_{\overline{\mathbf{B}}}$ is plotted ({\it dotted diagonal line}) to guide the eyes. The grey line represents the best linear fit.
{\it Right:} Result from Section~\ref{sec::clouds}; scatter plot of the difference between the most probable values of $p$-derived and 2D projected inclination angles, ${\gamma_\mathrm{obs}}^\wedge - {\gamma_\mathrm{2D}}^\wedge$, as a function of $\langle{\cal S}\rangle$, for both original ({\it smaller dots}) and ${\cal S}$-corrected ({\it larger dots}) inclination angles, color-coded by their corresponding ${\gamma_\mathrm{obs}}^\wedge$ values. 
Grey straight lines represent results from linear fitting, with thinner line representing fitting result using smaller dots (original ${\gamma_\mathrm{obs}}^\wedge$) and thicker line representing that from ${\cal S}$-corrected values (larger dots). 
The median value of ${\cal S}$ of Vela C \citep{2016ApJ...824..134F} is also included ({\it vertical dashed lines}, color-coded by its $p$-derived inclination angle; see Section~\ref{sec::velac}).
}
\label{fig::allrcomp}
\end{figure}

For each simulated cloud, we considered three different ways to calculate the representative inclination angle $\gamma$. The most straightforward way is to take the average of the magnetic field $\overline{\mathbf{B}}$ and use that direction as the mean inclination angle of the cloud:
\begin{align}
\overline{\mathbf{B}} & = \overline{B_x}~\hat{\mathbf{x}} + \overline{B_y}~\hat{\mathbf{y}} + \overline{B_z}~\hat{\mathbf{z}}, \label{eq::rB} \\
\gamma_{\overline{\mathbf{B}}} & \equiv \cos^{-1} \sqrt{\frac{\overline{B_x}^2 + \overline{B_y}^2}{|\overline{\mathbf{B}}|^2}}.
\label{eq::rB}
\end{align}
We can also directly calculate the inclination angle at each cell of the simulation box, $\gamma_\mathrm{3D}$:
\begin{equation}
\gamma_{\mathrm{3D},\ (i,j,k)} = \cos^{-1} \sqrt{\frac{{B_x}^2 + {B_y}^2}{|\mathbf{B}|^2}}\Bigg|_{(i,j,k)},
\label{eq::r3D}
\end{equation}
and use the most probable value (i.e.~the peak location of PDF), ${\gamma_\mathrm{3D}}^\wedge$, to represent the inclination angle of the cloud.

However, to be more comparable to the synthetic observations, which is projected onto 2D space, we define another quantity $\gamma_\mathrm{2D}$:
 \begin{equation}
\cos^2 \gamma_{\mathrm{2D},\ (i,j)} = \frac{\displaystyle\sum_{k} \rho_{i,j,k}\cdot \cos^2\gamma_{\mathrm{3D},\ (i,j,k)}}{\displaystyle\sum_{k} \rho_{i,j,k}}.
\label{eq::r2D}
\end{equation}
This is the density-weighted mean of inclination angle at each line of sight through the whole depth of the cloud. We again use the most probable value, ${\gamma_\mathrm{2D}}^\wedge$, to represent the cloud-scale inclination angle.

For all tests discussed in Sections~\ref{sec::clouds} and \ref{sec::rot}, we listed $\gamma_{\overline{\mathbf{B}}}$, ${\gamma_\mathrm{3D}}^\wedge$, and ${\gamma_\mathrm{2D}}^\wedge$ as comparisons to the $p$-derived inclination angle $\gamma_\mathrm{obs}$ (see Tables~\ref{simprop} and \ref{rotresult}). 
It is not surprising that ${\gamma_\mathrm{3D}}^\wedge$ could deviate from $\gamma_{\overline{\mathbf{B}}}$ by a certain amount, because mathematically the inclination angle cannot take into account the $\pm \gamma$ values; i.e.~antiparallel vectors will add constructively instead of cancel each other out.
This effect is more dramatic when the average magnetic field is around extreme values ($0^\circ$ and $90^\circ$), because by definition (Equation~(\ref{eq::r3D})) $\gamma_\mathrm{3D}$ is always within $[0^\circ, 90^\circ]$; for example, both vectors $10~\hat{\mathbf{x}} + \hat{\mathbf{z}}$ and $10~\hat{\mathbf{x}} - \hat{\mathbf{z}}$ would be considered to have inclination angle $\gamma \approx 6^\circ$, and thus the median value would be $6^\circ$ instead of $0^\circ$ as if directly calculated from averaging the vectors first.
The same happens to vectors $\hat{\mathbf{x}} + 10~\hat{\mathbf{z}}$ and $-\hat{\mathbf{x}} + 10~\hat{\mathbf{z}}$ around $\gamma=90^\circ$.
Similar errors also apply to the projected inclination angle $\gamma_\mathrm{2D}$, as well as the measured polarization angle in real observations. The $\pm \gamma$ for a given $|\gamma|$ will both give a positive contribution (of the same amplitude) to polarization, because intrinsically it is the projected ``shape'' of the spinning grains that determines the orientation of polarization.

Though $\gamma_{\overline{\mathbf{B}}}$ could in principle best represent the cloud-scale magnetic field direction, $\gamma_\mathrm{2D}$ is more practical when comparing with observations. 
We therefore plotted $\gamma_{\overline{\mathbf{B}}}$ vs. ${\gamma_\mathrm{2D}}^\wedge$ in Figure~\ref{fig::allrcomp} (left panel) using all synthetic observations discussed in the following sections; though the projected inclination angle ${\gamma_\mathrm{2D}}^\wedge$ has large errors with respect to the ``real" inclination angle $\gamma_{\overline{\mathbf{B}}}$ when $\gamma_{\overline{\mathbf{B}}} \sim 90^\circ$, there seems to be a nearly linear trend that could be used to estimate $\gamma_{\overline{\mathbf{B}}}$ from ${\gamma_\mathrm{2D}}^\wedge$. This could be useful when estimating the 3D magnetic field direction within individual MC from polarimetric observations, which we will discuss further in Section~\ref{sec::velac}.

\subsection{Test 1: Environmental Effect}
\label{sec::clouds}

\begin{figure}
\centering
\includegraphics[width=\columnwidth]{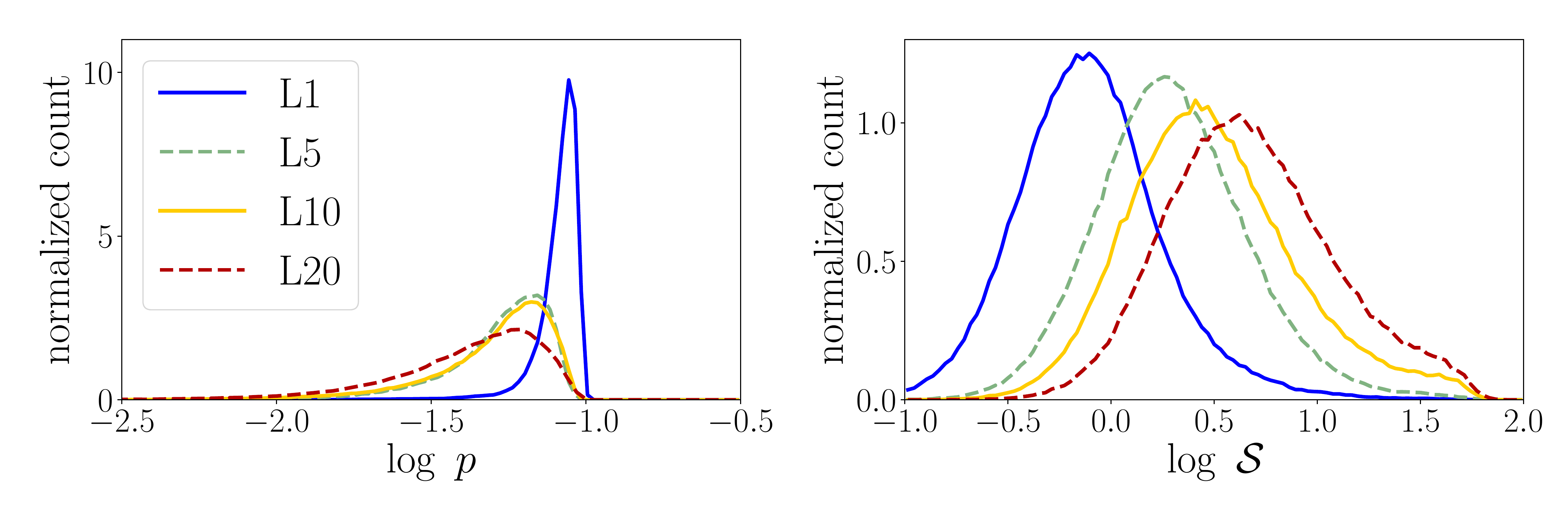}
\vspace{-.15in}
\caption{The PDFs of polarization fraction ({\it left}) and dispersion in polarization angles ({\it right}) from synthetic observations (along the $z$ direction) of the four simulation models considered in this study. Note that though the shapes of $p$ distributions vary significantly among models, the peak locations (the most probable values; see Table~\ref{simprop}) do not move far from the maximum value. In contrast, the shapes of ${\cal S}$ remain almost the same (as well as the widths; see Table~\ref{simprop} for the standard deviation $\sigma_{\log{\cal S}}$), regardless the very different peak locations (see the median values of ${\cal S}$ listed in Table~\ref{simprop}).}
\label{fig::model_pS}
\end{figure}

The four simulations listed in Table~\ref{simprop} have the same initial velocity perturbation pattern (see Figure~\ref{fig::cloudsim}), and have similar overall magnetic field direction $\gamma_{\overline{\mathbf{B}}}$ (mostly on the plane of sky; average inclination angle $\lesssim 15^\circ$). 
Though these models are all magnetically-dominated ($\beta_\mathrm{plasma} \lesssim 0.1-0.2$, which is common for shock-compressed regions), they are very different in terms of turbulence level ($v_\mathrm{rms}$; column 3 of Table~\ref{simprop}). This is reflected in the polarization level $p$ and the dispersion of polarization angle ${\cal S}$. Figure~\ref{fig::model_pS} compares the PDFs of $p$ and ${\cal S}$ for the four models; note that though the shape of $p$ distribution changes significantly from model L1 (the most quiescent) to model L20 (the most turbulent), the peak (i.e.~the most probable value of $p$, ${p}^\wedge$; see column 5 of Table~\ref{simprop}) does not change much. On the other hand, even though the median values of ${\cal S}$ are very different between models (a factor of $\gtrsim 5$; see column 7 of Table~\ref{simprop}), the shapes and widths of these PDFs remain similar (see column 8 of Table~\ref{simprop} for values of $\sigma_{\log {\cal S}}$, the standard deviation of ${\cal S}$ in $\log$ space).\footnote{Note that while we consider the peak location of PDF, or the most probable value, as the characteristic value for both polarization fraction and inclination angle, we use the median value of ${\cal S}$ to represent the level of polarization angle dispersion within the cloud. This is related to the fact that the PDFs of ${\cal S}$ being more symmetric than $p$ and $\gamma$, but also because when measuring ${\cal S}$ we are not trying to separate the background cloud (more pixels) from the overdense structures (smaller spatial coverage), and thus the median value is a good representative of the level of dispersion for the entire cloud.}

\begin{figure*}
\includegraphics[width=\textwidth]{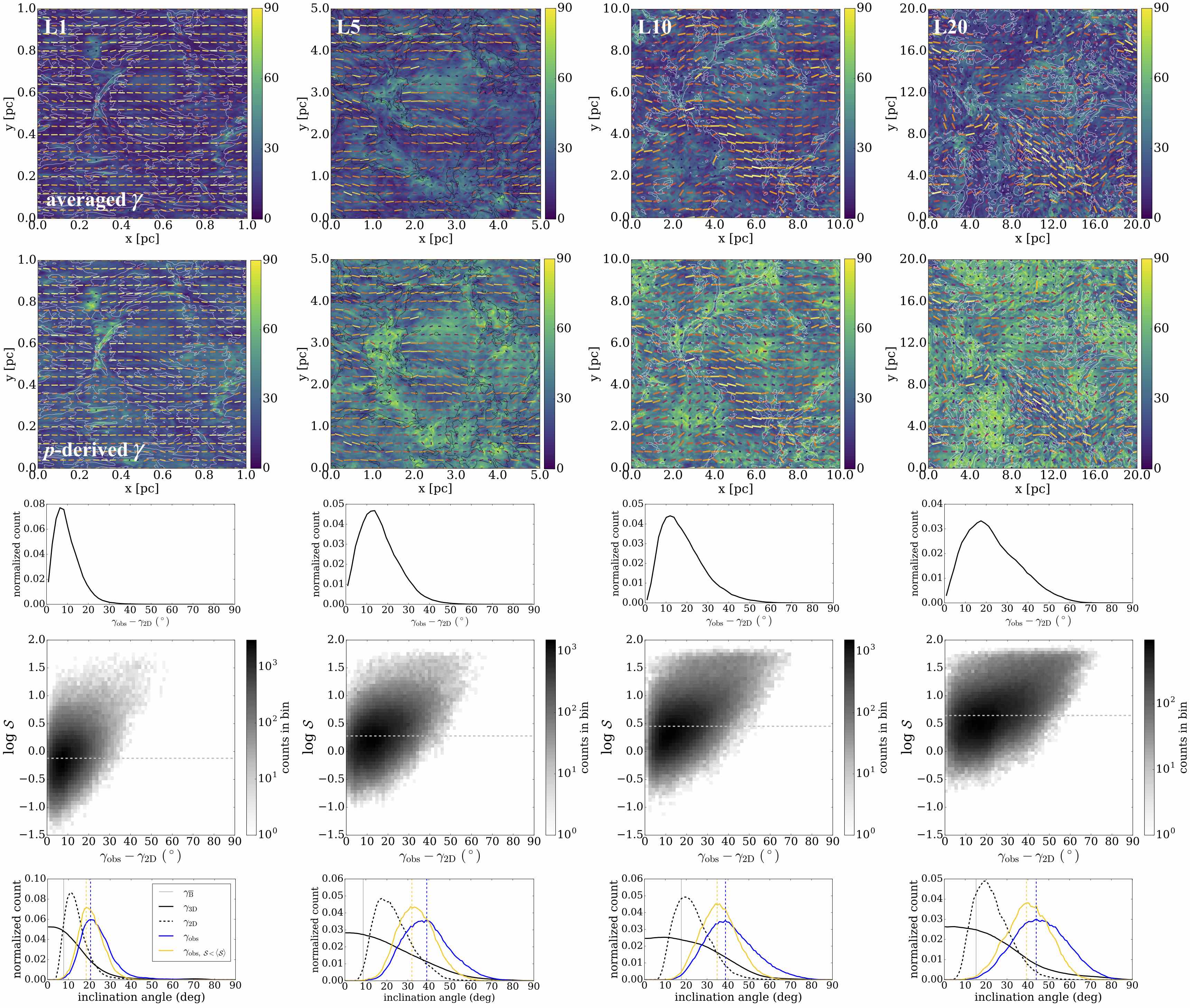}
\vspace{-.15in}
\caption{Results of the $p$-derived inclination angle from each model listed in Table~\ref{simprop} ({\it columns}). {\it Top row:} Maps of the projected inclination angle $\gamma_\mathrm{2D}$, with polarization vectors (color-coded by polarization fractions at the same colorscales as in the second row of Figure~\ref{fig::cloudsim}) and column density contours. {\it Second row:} Similar to the panels above, but showing $p$-derived inclination angle $\gamma_\mathrm{obs}$. Note that $\gamma_\mathrm{obs}$ almost follows the same pattern as $\gamma_\mathrm{2D}$, just at larger angles, because by theory $\gamma_\mathrm{obs}$ is the maximum-possible inclination angle at each location. {\it Third row:} Distributions of the difference between predicted and projected angles, $\gamma_\mathrm{obs} - \gamma_\mathrm{2D}$; the very few amount of negative values are not shown here. In all models, most of the sightlines have angle error less than $\sim 30^\circ$. {\it Fourth row:} 2D histograms (in $\log$ scale) of polarization angle dispersion ${\cal S}$ and the angle difference $\gamma_\mathrm{obs} - \gamma_\mathrm{2D}$, which suggest the $p$-derived inclination angle is more accurate for regions with lower dispersion. The median values of ${\cal S}$ are indicated by grey dashed lines. {\it Bottom row}: Comparisons of the theoretical ($\gamma_{\overline{\mathbf{B}}}$, $\gamma_\mathrm{3D}$, $\gamma_\mathrm{2D}$; {\it grey vertical line, black solid and dashed curves}) and estimated (the original and ${\cal S}$-corrected $\gamma_\mathrm{obs}$; {\it blue and yellow curves}) inclination angles, with dashed vertical lines showing the most probable values of corresponding $\gamma_\mathrm{obs}$.
}
\label{fig::simcomp}
\end{figure*}

We made synthetic polarimetric observations (viewing along the $z$ direction with $p_0 = 0.1$ at grid-size resolution, which is $L_\mathrm{box}/512$; we will discuss the effect of resolution in Section~\ref{sec::velac}) of these simulations following Equation~(\ref{eq::qu}), then applied Equations~(\ref{eq::p0}) and (\ref{eq::cosrobs}) to calculate the $p$-derived inclination angle, $\gamma_\mathrm{obs}$. The measured $p_\mathrm{max}$ and the correspondingly derived $p_0$, as well as the median value of $\gamma_\mathrm{obs}$, are listed in Table~\ref{simprop}.  
Figure~\ref{fig::simcomp} compares the density-weighted average inclination angle $\gamma_\mathrm{2D}$ (top row) and the $p$-derived $\gamma_\mathrm{obs}$ (second row) for each model; one can clearly see that the $p$-derived angle follows the same patterns as the 2D projected angle, but is generally larger. This is expected, since Equation~(\ref{eq::cosrobs}) considers the ``perfect'' scenario when there is no variation in position angle $\psi$ along the line of sight, and therefore $\gamma_\mathrm{obs}$ could be considered as the maximum possible $\gamma$ instead of the actual inclination angle. 

The difference between $\gamma_\mathrm{2D}$ and $\gamma_\mathrm{obs}$, or the ``error'' of the $p$-derived inclination angle, is thus defined as $\Delta\gamma \equiv \gamma_\mathrm{obs} - \gamma_\mathrm{2D}$ and is plotted as a PDF of all pixels from the synthetic observation map in Figure~\ref{fig::simcomp} (middle row). The error in estimated inclination angle is definitely dependent on the level of turbulence; for models with larger RMS velocity ($v_\mathrm{rms}$), there are about half of the pixels with errors $\Delta\gamma \gtrsim 20^\circ-30^\circ$ (models L10 and L20). In contrast, for the least perturbed model L1, almost all pixels have errors within $\sim 20^\circ$.

The correlation between the accuracy of the $p$-derived inclination angle and the turbulence level within the cloud can be inferred from the correlation between $\Delta\gamma$ and the dispersion in polarization angle ${\cal S}$, which is illustrated as a 2D histogram in Figure~\ref{fig::simcomp} (fourth row). It is obvious that pixels with smaller ${\cal S}$, in general, tend to have smaller errors in $p$-derived inclination angle; in fact, if we measure the most probable value of $\gamma_\mathrm{obs}$ among only the pixels with ${\cal S}$ less than its median value ($\langle{\cal S}\rangle$; grey dashed line in the 2D histogram), i.e.~considering only half of the pixels from the map that are less perturbed (the yellow-brown regions in the ${\cal S}$ maps in Figure~\ref{fig::cloudsim}), it could be different from the most probable value of the whole map and could be more accurate (closer to $\gamma_{\overline{\mathbf{B}}}$). These ${\cal S}$-corrected ${\gamma_\mathrm{obs}}^\wedge$ are also listed in Table~\ref{simprop}.

We summarize the results from this set of tests in the bottom row of Figure~\ref{fig::simcomp}. The cloud-scale inclination angle ($\gamma_{\overline{\mathbf{B}}}$, grey vertical lines), the averaged inclination angle in 3D ($\gamma_\mathrm{3D}$, black solid curves), the averaged projected inclination angle ($\gamma_\mathrm{2D}$, black dashed curves), and the $p$-derived inclination angle ($\gamma_\mathrm{obs}$, blue curves) are plotted together for comparison. 
Generally speaking, our method of deriving magnetic field direction provides estimates of inclination angles within $\sim 30^\circ$.
Also included in the bottom row of Figure~\ref{fig::simcomp} is the ${\cal S}$-corrected $\gamma_\mathrm{obs}$ (yellow curves), which only includes pixels with dispersions less than the median value $\langle{\cal S}\rangle$ of the entire map. The low-${\cal S}$-selected sightlines obviously give a better estimate for ${\gamma_\mathrm{obs}}^\wedge$, but only by a small amount ($\lesssim 5^\circ$). 
The most probable values (the location of the peak determined from Gaussian KDE) of both the original and ${\cal S}$-corrected $\gamma_\mathrm{obs}$ are also included as dashed vertical lines for easy comparisons.

A quantitative correlation between $\Delta\gamma^\wedge \equiv {\gamma_\mathrm{obs}}^\wedge - {\gamma_\mathrm{2D}}^\wedge$ and $\langle{\cal S}\rangle$ is shown in the right panel of Figure~\ref{fig::allrcomp}, for both the original (smaller dots) and ${\cal S}$-corrected (larger dots) ${\gamma_\mathrm{obs}}^\wedge$. Clearly, for models with similar inclination angles (indicated by colors), the error in $p$-derived inclination angle follows roughly a linear dependence on the level of dispersion of the cloud (grey lines). Though this could be useful when applying this method to estimate the magnetic field direction within a cloud, we note that the inclination angle itself could be an important factor determining the error of this method (see Section~\ref{sec::velac} for the case of Vela C). 
This is because the basic theory of $p$-derived inclination angle, which assumes no structure along each sightline, is intrinsically less applicable when the inclination angle is small; when most of the magnetic field is aligned with the plane of sky, the effect of depolarization is dominated by variation of $\psi$ along the line of sight, which is neglected in Equation~(\ref{eq::pobs}). It is not until the inclination angle is large enough to be responsible for most of the depolarization that our method would produce a more accurate prediction. 
Also, as discussed in Section~\ref{sec::defgamma}, the projected inclination angle intrinsically has larger error when the average magnetic field orientation is roughly either on the plane of sky ($\gamma_{\overline{\mathbf{B}}} \approx 0^\circ$) or along the line of sight ($\gamma_{\overline{\mathbf{B}}} \approx 90^\circ$).
This is the main focus of the following section below.

\subsection{Test 2: Viewing Angle}
\label{sec::rot}

\begin{table*}
	\centering
	\caption{Similar to Table~\ref{simprop}, but showing the results from different viewing angle of the same simulation (model L10).}
	\label{rotresult}
	\begin{tabular}{c | cc | c | ccccc} 
		\hline
		rotating & \multirow{2}{*}{${p}^\wedge$}  & \multirow{2}{*}{$\langle{\cal S}\rangle$} & derived & $\gamma_{\overline{\mathbf{B}}}$ & ${\gamma_\mathrm{3D}}^\wedge$ & ${\gamma_\mathrm{2D}}^\wedge$ & ${\gamma_\mathrm{obs}}^\wedge$ & ${\gamma_{\mathrm{obs}_,\ {\cal S}<\langle{\cal S}\rangle}}^\wedge$\\
		angle & &  & $p_0$ & $(^\circ)$ & $(^\circ)$ & $(^\circ)$ & $(^\circ)$ & $(^\circ)$\\
		\hline
		$\ \ 15^\circ$ & 0.073 & 2.8$^\circ$ & 0.095 & $\ \ 2.5$ & \ \ 1.3 & 15.6 & 32.6 & 29.6 \\
		$-15^\circ$ & 0.056 & 3.1$^\circ$ & 0.096 & $32.5$ & 33.1 & 28.0 & 45.9 & 41.4 \\
		$-45^\circ$ & 0.035 & 5.0$^\circ$ & 0.096 & $62.5$ & 55.6 & 46.4 & 58.9 & 53.5 \\
		$-75^\circ$ & 0.014 & 7.3$^\circ$ & 0.093 & $87.5$ & 67.2 & 57.1 & 69.2 & 67.5 \\
		\hline
		Vela C & 0.043 & 7.9$^\circ$ & 0.142 & $-$ & $-$ & $-$ & 64.9 & 54.5 \\
		\hline
	\end{tabular}
\end{table*}

\begin{figure}
\centering
\includegraphics[width=\columnwidth]{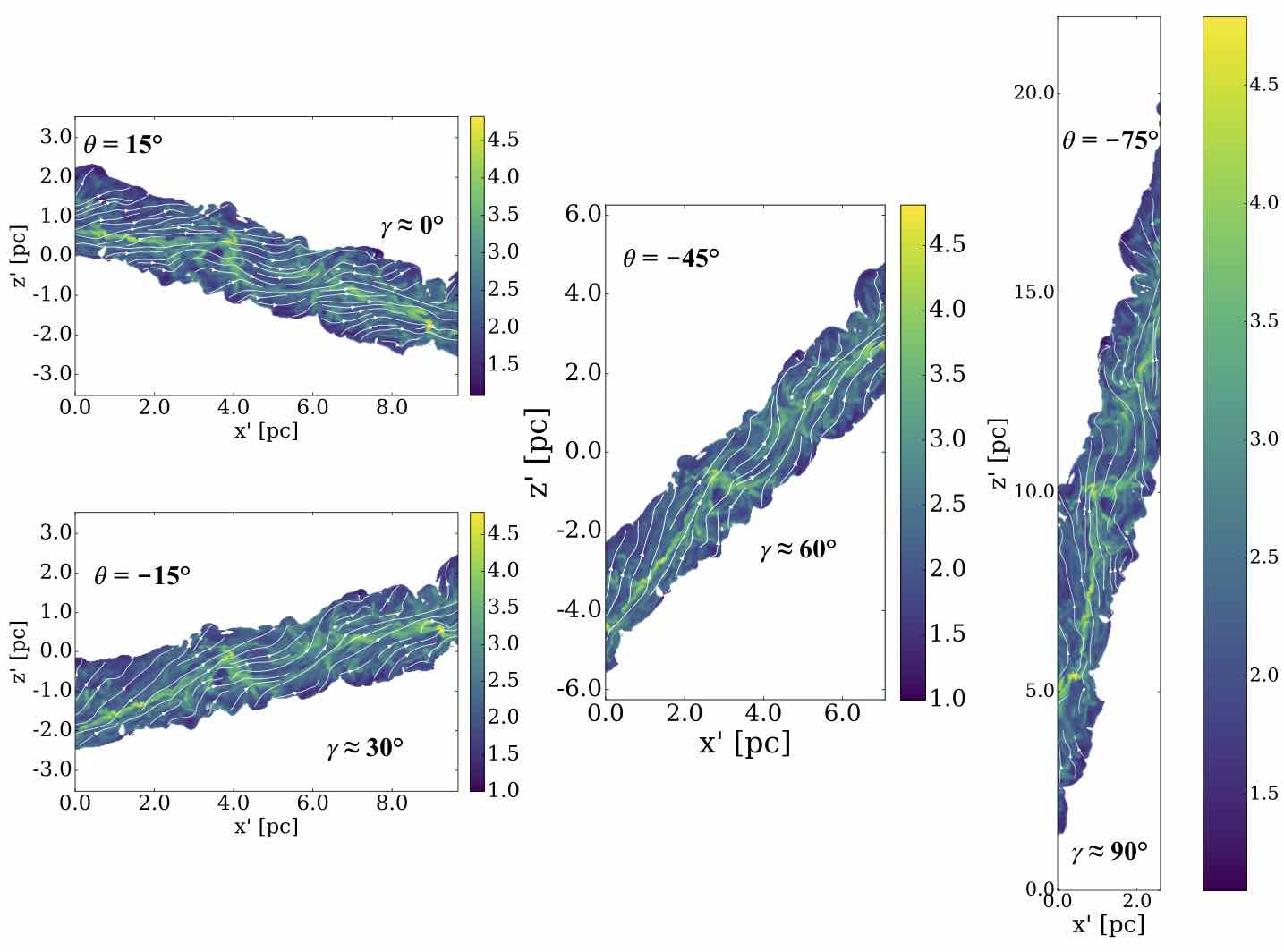}
\vspace{-.2in}
\caption{Number density of hydrogen nucleus $n_\mathrm{H}$ (colormaps, in $\log$-scale) and magnetic field structures (white streamlines) of slices of model L10, cut through $y\approx 0.4~L_\mathrm{box}$. The simulation box is rotated by various angles around $y$ axis to generate a range of cloud-scale magnetic field directions when viewed along the $z'$ direction. Note that only the post-shock regions are shown here.}
\label{fig::rotlayers}
\end{figure}

\begin{figure*}
\includegraphics[width=\textwidth]{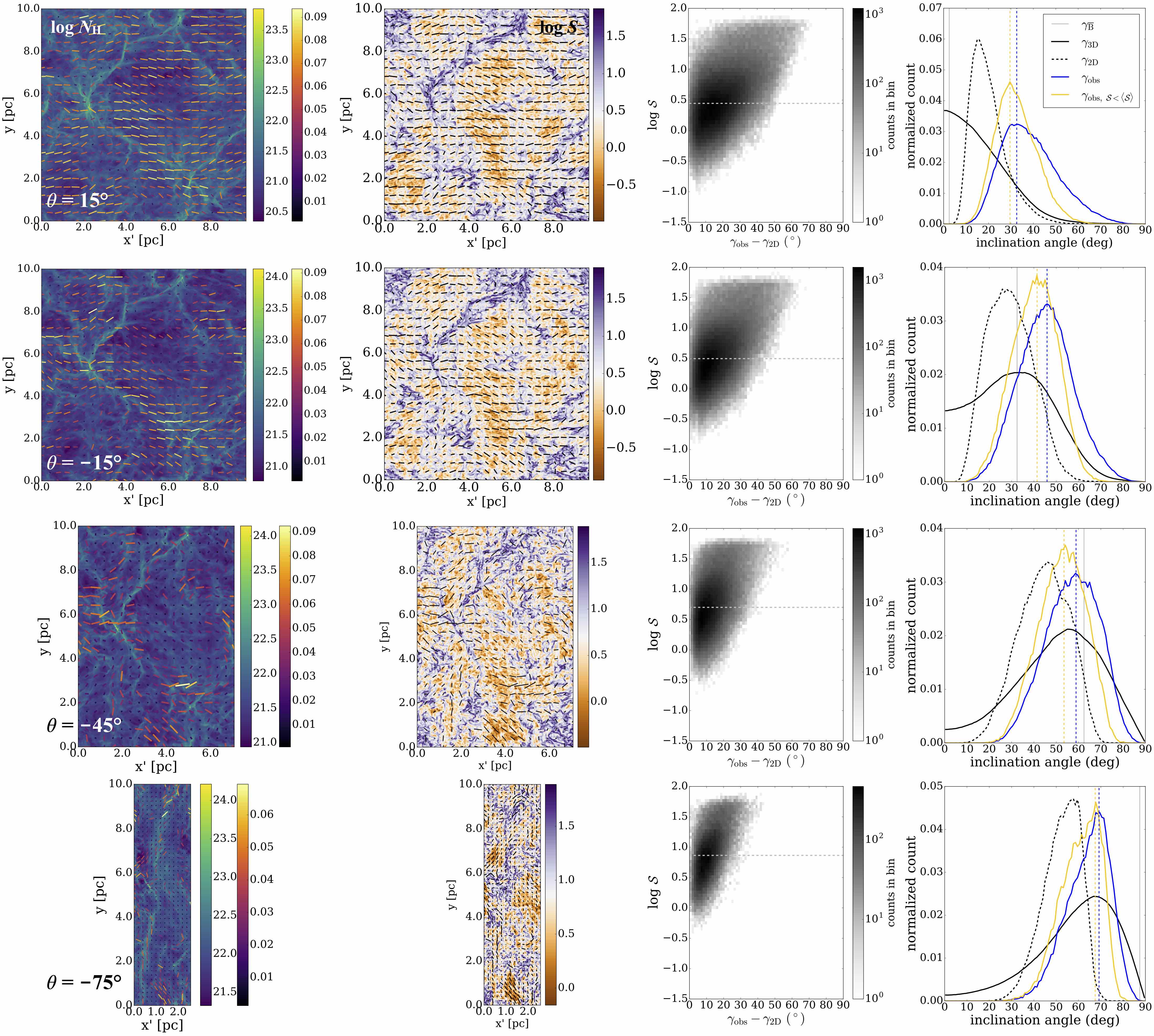}
\vspace{-.2in}
\caption{Results of synthetic observations from different viewing angles of model L10, as listed in Table~\ref{rotresult}. {\it First column:} synthetic observations of rotated models along the new $z$ direction, with column density $N_\mathrm{H}$ (colormap, in $\log$-scale) and polarization vectors (color-coded by polarization fractions). {\it Second column:} maps of dispersion of polarization angle, ${\cal S}$, with polarization vectors. The colorbars of these maps are centered at the median values of ${\cal S}$ so that regions with ${\cal S} < \langle{\cal S}\rangle$ are yellow-brown colored. {\it Third column:} 2D histograms (in $\log$ scale) of dispersion in polarization angles ${\cal S}$ and the angle difference $\gamma_\mathrm{obs} - \gamma_\mathrm{2D}$, with the median values of ${\cal S}$ indicated by grey dashed lines. The $p$-derived inclination angle clearly is more accurate when the magnetic field is more aligned with the line of sight. {\it Fourth column:} Comparisons of the theoretical ($\gamma_\mathrm{3D}$, $\gamma_\mathrm{2D}$; {\it black solid and dashed lines}) and estimated (the original and ${\cal S}$-corrected $\gamma_\mathrm{obs}$; {\it blue and yellow lines}) inclination angles, with dashed vertical lines showing the most probable values of corresponding $\gamma_\mathrm{obs}$. The inclination angles of the average magnetic field in 3D, $\gamma_{\overline{\mathbf{B}}}$, are also plotted as grey thin vertical lines for each viewing angle.
}
\label{fig::rotrcomp}
\end{figure*}

\begin{figure}
\centering
\includegraphics[width=\columnwidth]{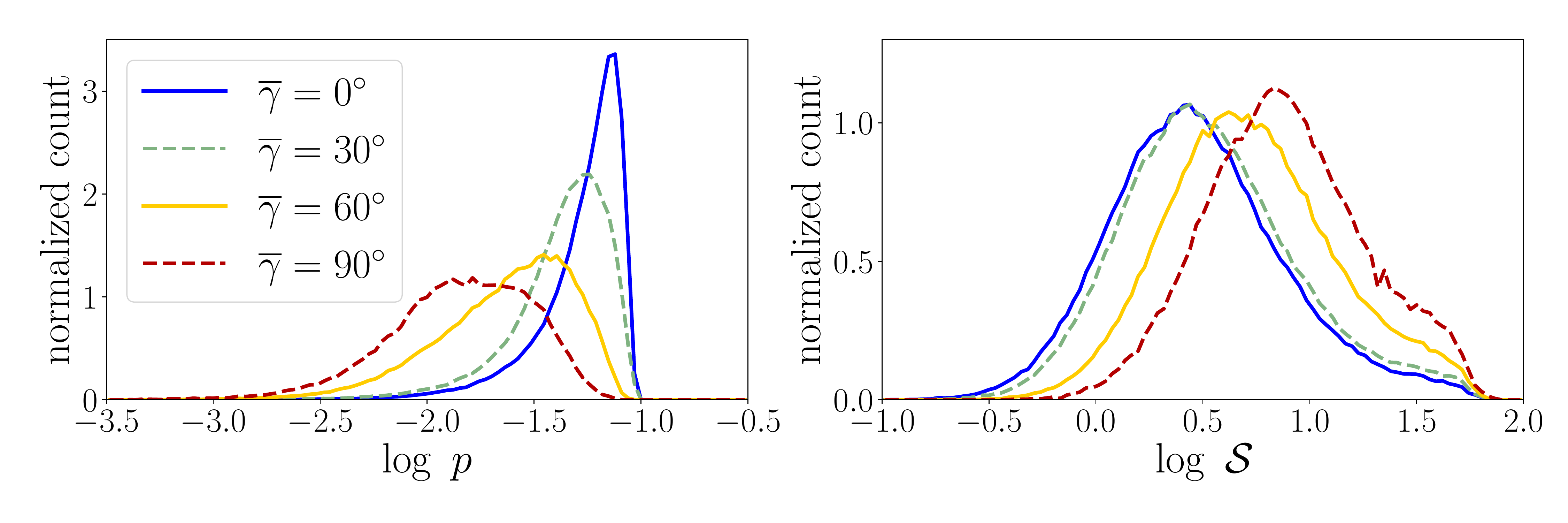}
\vspace{-.2in}
\caption{Similar to Figure~\ref{fig::model_pS}, but for different viewing angles of model L10. Note that the peaks of $p$ distribution changes dramatically for different viewing angles (i.e.~different inclination angle of the magnetic field).}
\label{fig::rotpScomp}
\end{figure}

\begin{figure}
\centering
\includegraphics[width=\columnwidth]{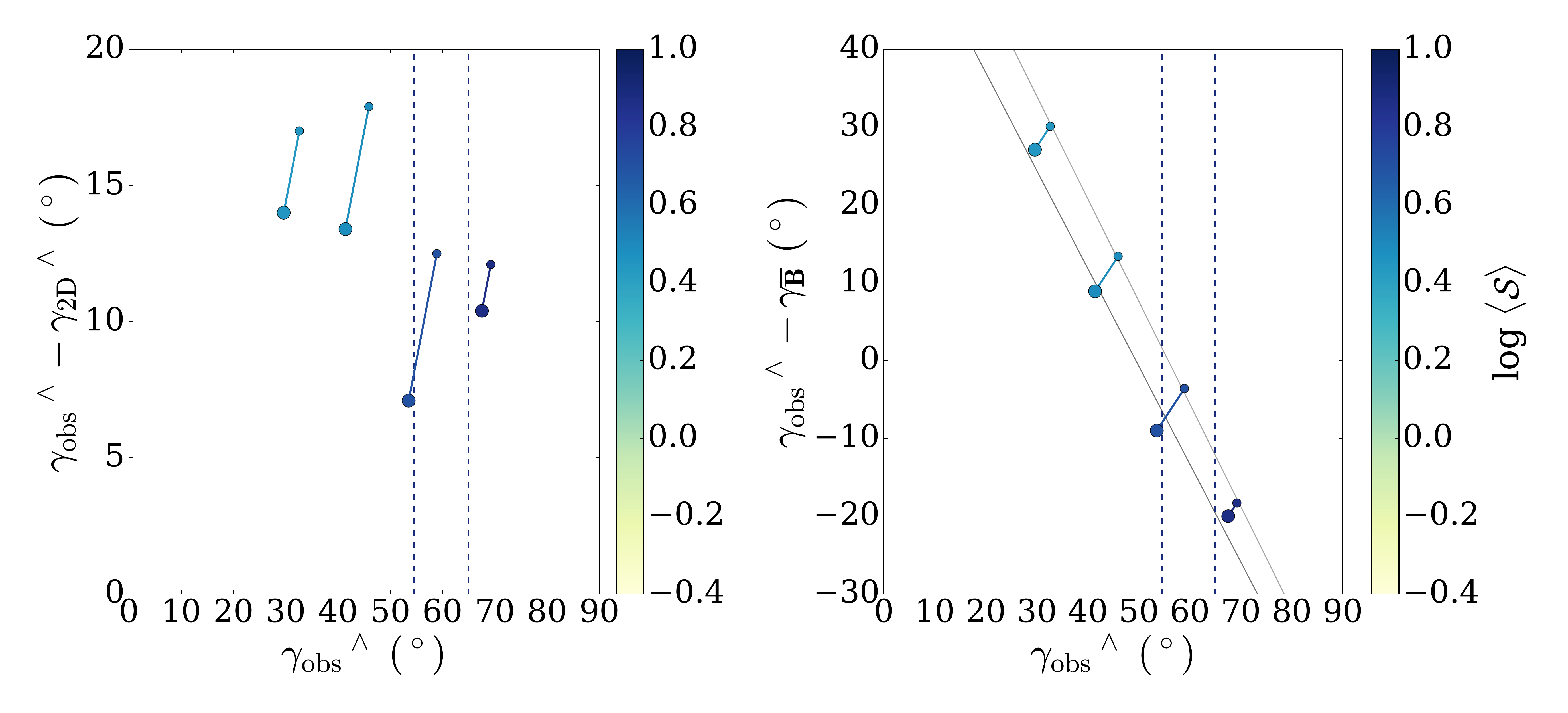}
\vspace{-.2in}
\caption{Results from Sections~\ref{sec::rot}; scatter plots of the difference between the most probable values of $p$-derived and 2D projected inclination angles (${\gamma_\mathrm{obs}}^\wedge - {\gamma_\mathrm{2D}}^\wedge$, {\it left}), and the difference between ${\gamma_\mathrm{obs}}^\wedge$ and cloud-scale inclination angle $\gamma_{\overline{\mathbf{B}}}$ ({\it right}),
as functions of ${\gamma_\mathrm{obs}}^\wedge$, for both original ({\it smaller dots}) and ${\cal S}$-corrected ({\it larger dots}) inclination angles, color-coded by their corresponding median ${\cal S}$ values (in $\log$ scale).
The grey straight lines in the right panel represent linear fitting results using the original ${\gamma_\mathrm{obs}}^\wedge$ (thinner line) and the ${\cal S}$-corrected values (thicker line). 
The results of Vela C (see Section~\ref{sec::velac}) using data from \protect\cite{2016ApJ...824..134F} are also included ({\it vertical dashed lines}, color-coded by the median value of ${\cal S}$ among Vela C).
}
\label{fig::rotr2DrB}
\end{figure}

Though the four simulations discussed in the previous section cover wide ranges of physical properties and dispersion levels, their average magnetic field orientations are similar and roughly on the plane of sky when viewed along the $z$-axis. 
We therefore selected model L10 as the test model, and rotated the simulation box to change the viewing angle so that the averaged magnetic field inclination angle varies between $\sim 0-90^\circ$. The corresponding rotating angles are listed in Table~\ref{rotresult}; since model L10 initially has mean magnetic field inclination angle $\gamma_{\overline{\mathbf{B}}} \sim 15^\circ$ (see Table~\ref{simprop} and Figure~\ref{fig::cloudsim}), we picked $\theta_\mathrm{rot} = 15^\circ$, $-15^\circ$, $-45^\circ$, and $-75^\circ$ around $+y$ direction to get $\gamma_{\overline{\mathbf{B}}} \sim 0^\circ$, $30^\circ$, $60^\circ$, and $90^\circ$. This is demonstrated in Figure~\ref{fig::rotlayers}, and the resulting maps of synthetic observation (column density, polarization, and polarization angle dispersion) from each viewing angle are illustrated in Figure~\ref{fig::rotrcomp} (first and second columns). 

Figure~\ref{fig::rotpScomp} contains the PDFs of polarization fraction and polarization angle dispersion. Compare to Figure~\ref{fig::model_pS}, we clearly see that in addition to varying shapes of $p$ distribution, the peak location of $p$ ($p^\wedge$; see Table~\ref{rotresult}) also changes accordingly to the inclination angle. This reflects the core concept of our method of deriving the inclination angle based on the variation of polarization fraction. The angle dispersion, however, does not move as dramatically as in Figure~\ref{fig::model_pS} for varying turbulence level. This indicates that the polarization angle dispersion is more directly related to the physical properties of the cloud rather than the viewing angle. We discuss further details about other possible controlling factors of $p$ and ${\cal S}$ in another study (King~et al., {\it in prep}).

Similar to Figure~\ref{fig::simcomp}, we investigate the correlation between polarization angle dispersion, ${\cal S}$, and the error in $p$-derived inclination angle, $\Delta\gamma\equiv \gamma_\mathrm{obs} - \gamma_\mathrm{2D}$, by plotting the 2D histogram in Figure~\ref{fig::rotrcomp} (third column). The range of the error is significantly reduced with larger inclination angle. 
In fact, after corrected by selecting only pixels with dispersions smaller than the median value $\langle{\cal S}\rangle$ (grey dashed lines in the 2D histogram), the $p$-derived inclination angle could be as accurate as within $\lesssim 10^\circ$ from ${\gamma_\mathrm{2D}}^\wedge$ (models of $\theta = -45^\circ$ and $-75^\circ$; see columns 6 and 8 of Table~\ref{rotresult}). 

The last column of Figure~\ref{fig::rotrcomp} summarizes and compares all theoretical and $p$-derived inclination angles from this set of data.
Intriguingly, we note that for models with larger inclination angles (models of $\theta = -45^\circ$ and $-75^\circ$; $\gamma \approx 60^\circ$ and $90^\circ$), the $p$-derived values follow the inclination angles averaged in 3D ($\gamma_\mathrm{3D}$, black curves), and therefore the theoretical, cloud-scale inclination angles ($\gamma_{\overline{\mathbf{B}}}$, grey vertical lines), better than the 2D projected values ($\gamma_\mathrm{2D}$, dashed black curves).
 As we already know there is intrinsic error between the 2D projected and the cloud-scale inclination angles (see Figure~\ref{fig::allrcomp}, left panel), these results suggest that the $p$-derived inclination angle could be more accurately tracing $\gamma_{\overline{\mathbf{B}}}$ than the density-weighted mean, $\gamma_\mathrm{2D}$ under certain circumstances, e.g.~when the inclination angle is large and is the dominant effect controlling the depolarization of the cloud.
We will make use of this result when applying this method to observational data of real clouds (see Section~\ref{sec::velac}).

We again quantitatively investigate the dependence of angle difference $\Delta\gamma \equiv \gamma_\mathrm{obs} - \gamma_\mathrm{2D}$ on the inclination angle by scatter-plotting their most probable values in Figure~\ref{fig::rotr2DrB} (left panel) for both original (smaller dots) and ${\cal S}$-corrected (larger dots) $\gamma_\mathrm{obs}$. 
Though we clearly see the error is reduced with larger inclination angle regardless of the increasing dispersion level (indicated by colors), the change is not monotonic, and therefore a simple linear fit would not be appropriate here as applied in Section~\ref{sec::clouds}.
This is indeed expected, because (as discussed in Section~\ref{sec::defgamma}) 
the projected angle $\gamma_\mathrm{2D}$ could introduce further errors with respect to the cloud-scale value when the inclination angle is either small ($\lesssim 10^\circ$) or large ($\gtrsim 60^\circ$; see the left panel of Figure~\ref{fig::allrcomp}). We therefore directly compare the $p$-derived values to the cloud-scale inclination angle, and plot ${\gamma_\mathrm{obs}}^\wedge - \gamma_{\overline{\mathbf{B}}}$ in the right panel of Figure~\ref{fig::rotr2DrB}. The contamination from $\gamma_\mathrm{2D}$ is now removed, and the angle difference ${\gamma_\mathrm{obs}}^\wedge - \gamma_{\overline{\mathbf{B}}}$ for both the original and ${\cal S}$-corrected ${\gamma_\mathrm{obs}}^\wedge$ values seem to nicely follow linear correlations with ${\gamma_\mathrm{obs}}^\wedge$ (grey straight lines). 
These promising results provide the basis for applying this method to real clouds, which we discuss in the following section.

\section{Application: the Vela C Molecular Cloud}
\label{sec::velac}

\begin{figure}
\centering
\includegraphics[width=\columnwidth]{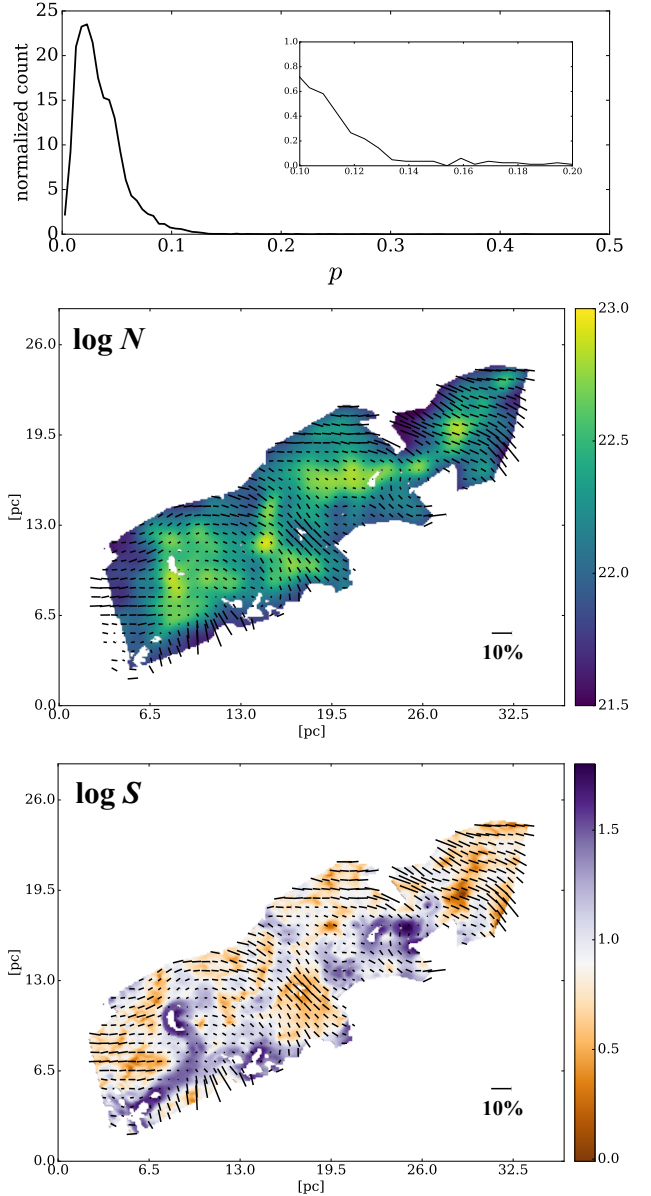}
\vspace{-.15in}
\caption{The polarimetric observation of Vela C, adopted from \protect\cite{2016ApJ...824..134F}. {\it Top:} PDF of polarization fraction measured in the entire Vela C cloud, zoomed-in to $p<0.2$ in the inset, which is used to determine $p_\mathrm{max}=0.15$. {\it Middle:} the map of Vela C in column density with polarization vectors (color-coded by polarization fraction). {\it Bottom:} dispersion of polarization angles in Vela C (in log scale). The colorbar is centered at the median value of ${\cal S}$ so that regions with ${\cal S} < \langle{\cal S}\rangle$ are yellow-brown colored.}
\label{fig::velacmap}
\end{figure}

\begin{figure}
\centering
\includegraphics[width=\columnwidth]{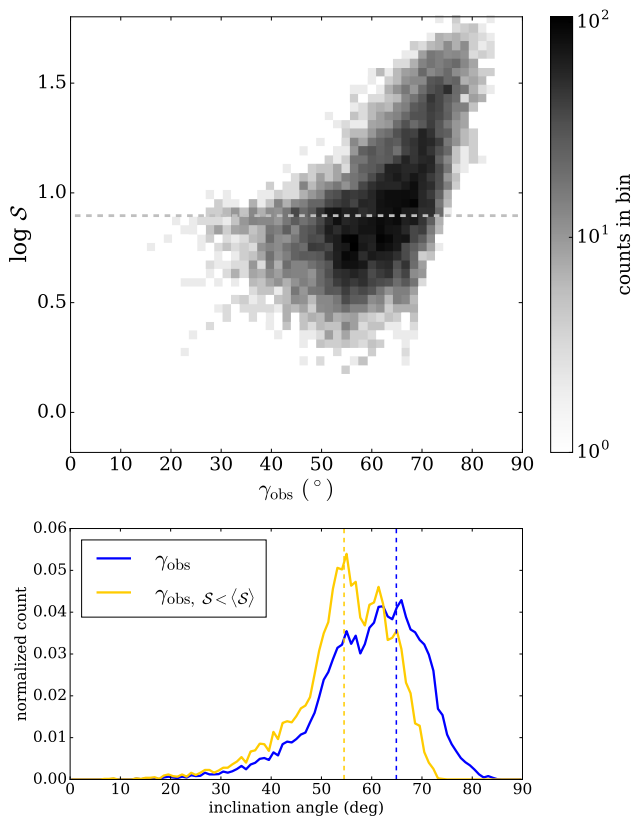}
\vspace{-.15in}
\caption{The estimated magnetic field inclination angle of Vela C. {\it Top:} 2D histogram of $\log {\cal S}$ and $p$-derived $\gamma_\mathrm{obs}$, with grey dashed line showing the median value of ${\cal S}$. {\it Bottom:} PDFs of $p$-derived $\gamma_\mathrm{obs}$, both original ({\it blue}) and ${\cal S}$-corrected ({\it yellow}). Dashed lines indicate the most probable values ($64.9^\circ$ and $54.5^\circ$) for both distributions.
}
\label{fig::velacrobs}
\end{figure}

We applied our method to 500 $\mu$m polarization observations of the Vela\,C giant molecular cloud using data from the BLASTPol telescope during its Antarctic flight of 2012--2013.  Vela\,C has a mass of approximately $\sim 10^5$~M$_\odot$ \citep{1999PASJ...51..775Y}, is relatively nearby (d\,=\,700\,$\pm$\,200\,pc; \citealt{1992A&A...265..577L}), and is generally cold ($T\,\leq\,$15\,K) with the exception of one region where two O9.5 stars are powering a compact bipolar \textsc{Hii} region \citep{2009ApJ...707.1824N, 2011A&A...533A..94H, 2013A&A...558A.102E}. Since Vela C is the most detailed polarimetric map yet made of a star-forming MC (with $>10^3$~independent polarization measurements made over the $\sim$30~pc cloud), it is an excellent target for statistical studies such as \hyperlink{KFCL18}{KFCL18} and our proposed method here.

To ensure that the detected polarized emission is from Vela C instead of background/foreground diffuse ISM dust, \cite{2016ApJ...824..134F} used different methods to identify and subtract diffuse polarized emission. Here we consider only their ``intermediate'' method of diffuse emission subtraction, however, we have verified that the results do not change significantly for different diffuse emission separation methods.  We also only include dust sightlines within the four Vela C cloud subregions defined in \cite{2011A&A...533A..94H}. To avoid errors introduced by de-biasing the polarization data, we only include sightlines with polarization signal-to-noise levels > 5 (see \cite{2016ApJ...824..134F} for more discussions). A polarization map of Vela C is shown in Figure~\ref{fig::velacmap} (middle and bottom panels).

To determine the best value of $p_\mathrm{max}$, we examined the PDF of $p$ from the whole map (Figure~\ref{fig::velacmap}, top panel), then picked the value that is on the edge between the Gaussian-like curve and the flat long tail. After zooming in (see the inset of the top panel of Figure~\ref{fig::velacmap}), we picked $p_\mathrm{max} = 0.15$ for the entire Vela C Cloud.
Note that the choice of $p_\mathrm{max}$ would definitely affect the derived values of $p_0$ and $\gamma_\mathrm{obs}$; however, considering the intrinsic error between the projected ($\gamma_\mathrm{2D}$) and the real-space magnetic field inclination angle ($\gamma_{\overline{\mathbf{B}}}$; see Figure~\ref{fig::allrcomp}, left), the error from $p_\mathrm{max}$ is relatively insignificant in comparison. 
Figure~\ref{fig::velacrobs} shows the PDF of $p$-derived inclination angle ($\gamma_\mathrm{obs}$, bottom panel) as well as the correlation between the dispersion of polarization angle ${\cal S}$ and $\gamma_\mathrm{obs}$ as a 2D histogram (top panel). The most probable values of $p$-inferred inclination angles (i.e.~the location of the peak measured from Gaussian KDE), both original ($64.9^\circ$) and ${\cal S}$-corrected ($54.5^\circ$), are listed in both Tables~\ref{simprop} and \ref{rotresult}. 

We first refer to Figure~\ref{fig::allrcomp} to investigate the accuracy of these $p$-derived inclination angle estimates. The right panel of Figure~\ref{fig::allrcomp} shows the scatter plot of the angle difference between $p$-derived and the 2D projected values, $\Delta\gamma\equiv {\gamma_\mathrm{obs}}^\wedge - {\gamma_\mathrm{2D}}^\wedge$, as a function of $\langle{\cal S}\rangle$, using the set of models discussed in Sections~\ref{sec::clouds}.
Though the linear fitting result from $\Delta\gamma$ vs.~$\langle{\cal S}\rangle$ indicates a very large error in ${\gamma_\mathrm{obs}}^\wedge$ of Vela C ($\sim 30^\circ$), we note that the inclination angle itself could be a critical factor in estimating the error of ${\gamma_\mathrm{obs}}^\wedge$ (see Section~\ref{sec::rot}). 
This is illustrated in the left panel of Figure~\ref{fig::rotr2DrB}; considering the result from this set of models with varying inclination angles, the $p$-derived ${\gamma_\mathrm{obs}}^\wedge$ of Vela C could be only $\sim 10^\circ$ higher than the actual 2D projected inclination angle. 
Roughly speaking, this suggests $\gamma_\mathrm{2D} \sim 45^\circ$ within Vela C (since the $p$-derived $\gamma$ is the maximum-possible inclination angle, theoretically $\gamma_\mathrm{obs}\geq\gamma_\mathrm{2D}$ should always hold).
Considering the discrepancy between the 2D projected value and the cloud-scale inclination angle,
if ${\gamma_\mathrm{2D}}^\wedge \approx 45^\circ$ for Vela C, our linear fit in Figure~\ref{fig::allrcomp} (left panel) indicates that the actual inclination angle of the magnetic field within Vela C is $\sim 65^\circ$.

On the other hand, from Section~\ref{sec::rot} we find that when the average inclination angle of the cloud-scale magnetic field is not small ($\gtrsim 30^\circ$), it is possible to directly link the $p$-derived inclination angle to the cloud-scale value, as demonstrated in the left panel of Figure~\ref{fig::rotr2DrB}. Therefore, by comparing the measured inclination angle of Vela C and the linear fit using those from simulations (also, note that the ${\cal S}$-corrected ${\gamma_\mathrm{obs}}^\wedge$ derived in Vela C is very close to that measured in our model $\theta = -45^\circ$, or $\gamma = 60^\circ$; see Table~\ref{rotresult}), our results suggest that the magnetic field in Vela C has an inclination angle $\gamma_{\overline{\mathbf{B}}} \sim 60^\circ$.

These values roughly agree with the results reported in \hyperlink{KFCL18}{KFCL18} using 2D correlations between polarization fraction, dispersion in polarization angles, and column density of the Vela C cloud and comparing with simulation models. 
Though the method adopted in \hyperlink{KFCL18}{KFCL18} is not completely independent from the analysis reported in this work (both methods consider the distribution of polarization fraction as a probe to magnetic field structure), the reliability of our new method is reassured by the consistency between these two studies. 
Therefore, it is likely that the cloud-scale magnetic field in Vela C has a large inclination angle, with a large line-of-sight component and only a relatively weak field on the plane of sky. However, we would like to point out that Vela C is a giant molecular cloud, which may contain regions with different magnetic field structures \citep[see e.g.~][]{2017A&A...603A..64S,2018arXiv180408979F}. As a result, the $p$-derived inclination angle may differ between regions. We will discuss this topic in a future study.
Another caveat is that we assumed that, on the scale of MCs, the variation of the polarization is dominated by the variation of the magnetic field configuration rather than the grain alignment efficiency. The latter is explored in detail in King et al. {\it (in prep.)}. 

We would also like to note that spatial resolution could have crucial impact on polarization measurements, especially the level of dispersion in polarization angles (see e.g.~Figure~3 of \hyperlink{KFCL18}{KFCL18}). In this study, we considered grid-size resolution for all of our synthetic observations, which are $L_\mathrm{box}/512 \approx 0.002$, 0.001, 0.02, 0.04~pc for models L1, L5, L10, L20, respectively. Though these isothermal, ideal-MHD simulations are free to be re-scaled (see discussions in \hyperlink{KFCL18}{KFCL18}), we note here that the spatial resolutions adopted in our synthetic observations are very different from the beam size of BLASTPol ($\sim 0.5$~pc in Vela C), and therefore may introduce further uncertainties when we translate $\gamma_\mathrm{obs}$ of Vela C into $\gamma_{\overline{\mathbf{B}}}$ or $\gamma_\mathrm{3D}$ using the results from these synthetic observations.

\section{Summary}
\label{sec::sum}

We propose a new method to estimate the direction of MC-scale magnetic field with respect to the plane of sky using dust polarization level measured in the cloud. By considering only regions with relatively low dispersion in polarization angles (where the depolarization is mostly determined by inclination angle), we successfully showed that this new method gives estimates for the magnetic field inclination angle with accuracy $\lesssim 10-30^\circ$. 
This method is further tested on BLASTPol polarimetric observation towards the Vela C Molecular Cloud, which suggests that the cloud-scale magnetic field direction in Vela C is inclined $\sim 60^\circ - 65^\circ$ away from the plane of sky. 

We summarize our main conclusions below:
\begin{enumerate}
\item Under the assumption of a uni-directional magnetic field along individual lines of sight, the basic equations for deriving dust polarization (Equations~(\ref{eq::qu})$-$(\ref{eq::polN2})) give a simple expression of the inclination angle of magnetic field $\gamma$ for each observed polarization fraction $p_\mathrm{obs}$ (Equation~(\ref{eq::cosrobs})). Since any variation of the plane-of-sky magnetic field position angle $\psi$ along the line of sight will further reduce the polarization level in addition to that imposed by inclination angle, this inferred value of $\gamma$ only represents the upper limit of inclination angle of magnetic field at a given location.

\item Assuming that the maximum polarization fraction within the cloud $p_\mathrm{max}$ can be measured (which only happens when there is at least one line of sight contains only uniform magnetic field on the plane of sky), the polarization coefficient $p_0$ can be directly derived from this maximum polarization fraction (Equation~(\ref{eq::p0})). Once $p_0$ is derived, every measured polarization fraction value can be used to infer the local, averaged inclination angle of magnetic field at that particular line of sight using Equation~(\ref{eq::cosrobs}).

\item We found that, using 3D MHD MC-scale simulations, the projected inclination angle ($\gamma_\mathrm{2D}$; Equation~(\ref{eq::r2D})) could differ from the averaged inclination angles calculated in 3D space ($\gamma_{\overline{\mathbf{B}}}$ and $\gamma_\mathrm{3D}$; Equations~(\ref{eq::rB}) and (\ref{eq::r3D})), especially when the 3D inclination angle is close to extreme values ($0^\circ$ or $90^\circ$; see the left panel of Figure~\ref{fig::allrcomp}). 
Nevertheless, under the criterion of similar dispersion level ${\cal S}$, the 3D inclination angle can be inferred from $\gamma_\mathrm{2D}$ using linear fitting from simulation data (see the left panel of Figure~\ref{fig::allrcomp}).

\item Based on previous work (\hyperlink{CKL16}{CKL16}; \hyperlink{KFCL18}{KFCL18}), the major depolarization effects within MCs (in addition to dust grain alignment efficiency) are the relative strength of line-of-sight magnetic field (i.e.~the inclination angle), and the dispersion level of the plane-of-sky magnetic field direction. This is further confirmed in our Monte Carlo experiments of measuring polarization fraction among models with different inclination angles and concentration levels of plane-of-sky magnetic field direction (Section~\ref{sec::MC}). We found that the polarization fraction$-$inclination angle correlation in our less-perturbed Monte Carlo models agrees quantitatively better with the theoretical prediction (Figure~\ref{MCplot}), which is the analytic solution for a perfect scenario (uniform magnetic field at each line of sight).

\item We examined the accuracy of the $p$-derived inclination angle in simulated clouds with different dispersion levels of polarization angle (Section~\ref{sec::clouds}), and different mean inclination angles of magnetic field (Section~\ref{sec::rot}). We confirmed that the $p$-derived inclination angle is more accurate in clouds with smaller ${\cal S}$; in addition, we found that by excluding regions with ${\cal S}$ larger than the median value $\langle{\cal S}\rangle$, the error is reduced in these ${\cal S}$-corrected values (see the yellow and blue curves in the bottom row of Figure~\ref{fig::simcomp}).
We also noted that our method of predicting $\gamma$ from $p$ is more applicable when the real $\gamma$ is away from the extreme values ($0^\circ$ and $90^\circ$), as the error ${\gamma_{\mathrm{obs}, {\cal S} < \langle{\cal S}\rangle}}^\wedge - \gamma_{\overline{\mathbf{B}}}$ is within $\sim \pm 10^\circ$ for simulated clouds with inclination angle $30^\circ-60^\circ$ (Figure~\ref{fig::rotrcomp}, right panel). 

\item We tested our method on the Vela C Molecular Cloud using polarimetric data from BLASTPol \citep{2016ApJ...824..134F}. The results from simulations (Section~\ref{sec::test}) are adopted to correct the errors in the $p$-derived inclination angles (Figures~\ref{fig::allrcomp} and \ref{fig::rotr2DrB}). Though there are still uncertainties in determining $p_\mathrm{max}$ (and hence $p_0$; see Table~\ref{simprop} for values), our result suggests that the cloud-scale magnetic field of Vela C is at $\sim 60^\circ - 65^\circ$ with respect to the plane of sky (see Figure~\ref{fig::velacrobs}), i.e.~much closer to the line of sight than to the plane of sky. This is in good agreement with the conclusion of \hyperlink{KFCL18}{KFCL18}. Considering Vela C contains sub-regions with very different physical properties \citep{2017A&A...603A..64S,2018arXiv180408979F}, future studies using the same method on individual sub-regions might be helpful to further improve the understanding of the magnetic field structure in Vela C.

\end{enumerate}

\section*{Acknowledgements}

We thank the referee for a very helpful report.
C.-Y.C. is grateful for the support from Virginia Institute of Theoretical Astronomy (VITA) at the University of Virginia through the VITA Postdoctoral Prize Fellowship. C.-Y.C., L.M.F., and Z.-Y. L. acknowledge support from NSF grant AST1815784.
Z.-Y.L. is supported in part by NSF AST1716259, and NASA NNX14AB38G and 80NSSC18K1095.
P.K.K. is supported by a Livermore Graduate Scholarship at Lawrence 
Livermore National Laboratory, and acknowledges additional support from ALMA SOS, as well as the Jefferson Scholars Foundation through a graduate fellowship. Part of this work was performed under the auspices of the Department of Energy by Lawrence Livermore National Laboratory under Contract DE-AC52-07NA27344. LLNL-JRNL-758117.
L.M.F. is a Jansky Fellow of NRAO. NRAO is a facility of the National Science Foundation (NSF operated under cooperative agreement by Associated Universities, Inc). 
The BLAST project is supported by NASA 80NSSC18K0481. 








\appendix

\section{Accuracy of Estimated \lowercase{$p_0$} and \lowercase{$\gamma_\mathrm{obs}$}}
\label{sec::appx}

Indeed, there are limits on the method of estimating the average inclination angle of cloud-scale magnetic field discussed in this manuscript. In principle, the measure of $p_\mathrm{max}$ (or equivalently, $p_0$) assumes that one of the observed sightlines has $\gamma \approx 0$ no matter what the cloud-scale inclination angle is. If the cloud-scale inclination angle is large (i.e.~closer to the line of sight), this could only happen when the dispersion of magnetic field direction is large, which means the cloud is strongly perturbed, or equivalently, has high Alfv{\'e}n Mach number ${\cal M}_\mathrm{A}$. However, under those conditions, depolarization due to the projection effect (cancellation among various position angles along the line of sight) may not be negligible, which may affect the accuracy of Equation~(\ref{eq::cosrobs}). 
To test how good $p_\mathrm{max}$ and $\gamma_\mathrm{obs}$ are recovered in various cloud conditions with different inclination angles, we expand our analysis in Section~\ref{sec::rot} to include models L1 (less perturbed) and L20 (most perturbed), and compare the results with those discussed in Section~\ref{sec::rot} using model L10 (moderately perturbed). 

\begin{figure}
\centering
\includegraphics[width=\columnwidth]{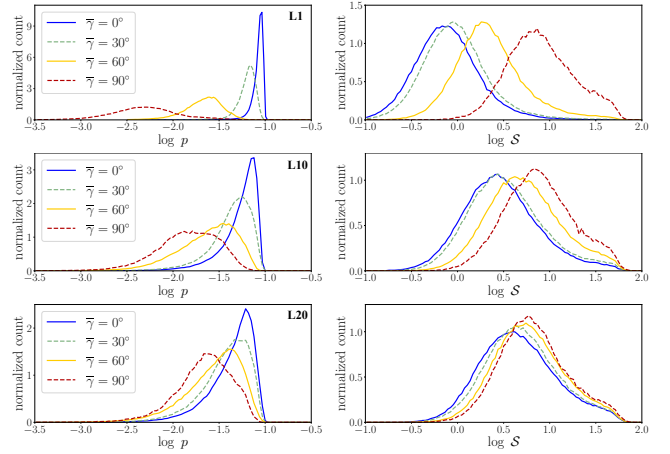}
\vspace{-.2in}
\caption{Similar to Figure~\ref{fig::rotpScomp}, but now comparing model L10 ({\it middle panels}) with two additional models, L1 ({\it top panels}) and L20 ({\it bottom panels}).}
\label{fig::rotpScompL20}
\end{figure}

These results are summarized in Figure~\ref{fig::rotpScompL20}, where we show the PDFs of polarization fraction $p$ and angle dispersion ${\cal S}$ for all three models, with four different inclination angles of the cloud-scale magnetic field. One can clearly see that the shape of the $p$ distribution (left panel of Figure~\ref{fig::rotpScompL20}) changes accordingly with inclination angle more dramatically in model L1 than models L10 and L20. This is because in less-perturbed environment, the dominant de-polarization effect comes from the inclination angle, while in more turbulent cloud, it is the projection effect that causes most of the depolarization. The ${\cal S}$ distributions in Figure~\ref{fig::rotpScompL20} (right panel) provide further evidence to this argument, because ${\cal S}$ is a measurement directly related to the projection effect.\footnote{Though the projection effect depends on the angle dispersion along the line of sight while ${\cal S}$ measures the angle dispersion on the plane of sky, these two quantities are closely correlated, as shown in Figure~13 of \hyperlink{CKL16}{CKL16}.} The fact that the peak locations of ${\cal S}$ remain almost the same under varying inclination angles in model L20 but move up significantly with increasing $\gamma$ in model L1 suggest that in less perturbed environment (L1), the projection effect is critical only when the inclination angle is large, while in more turbulent cloud like L20, depolarization caused by projection effect is crucial even when the cloud-scale magnetic field is closer to the plane of sky.

The quantitative measurements are listed in Table~\ref{rotL1L10L20}. As we expected, in more turbulent model L20, the value of $p_0$ is better recovered (comparing to the theoretical value $0.1$ adopted in our synthetic calculation) even when the cloud-scale inclination angle is large. In contrast, when the cloud is less perturbed and the magnetic field is more organized (model L1), it is almost impossible to get a correct estimate of $p_0$ when the inclination angle is large, due to the absence of sightlines with $\gamma \approx 0$. However, we also see that the accuracy of $p$-derived inclination angle does not seem to be affected much by the uncertainty of $p_0$, as model L1 still gives better estimate of the cloud-scale inclination angle ($\gamma_{\mathrm{obs}_,\ {\cal S}<\langle{\cal S}\rangle}$ closer to $\gamma_{\overline{\mathbf{B}}}$) even in highly inclined cases. This suggest that our method, Equation~(\ref{eq::cosrobs}), works nicely in less-perturbed clouds even when the magnetic field is almost aligned with the observer's line of sight. More importantly, this implies that the cloud-scale inclination angle is better reflected by the difference between $p_\mathrm{max}$ and $p^\wedge$, not the absolute values of those quantities, which is consistent with our method of deriving $\gamma$ from $p$. 

\begin{figure}
\centering
\includegraphics[width=\columnwidth]{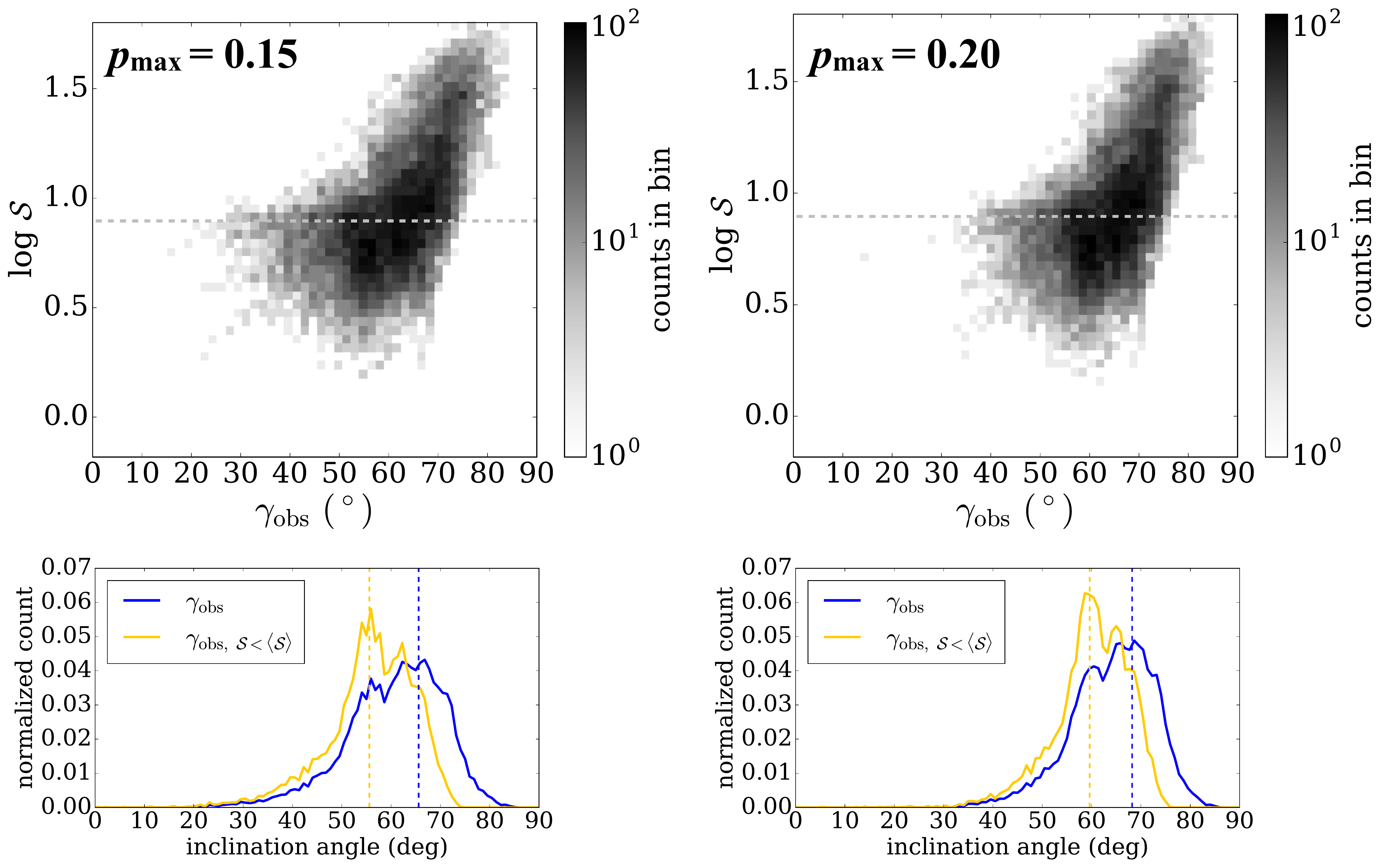}
\vspace{-.15in}
\caption{Similar to Figure~\ref{fig::velacrobs}, but now comparing two $p_\mathrm{max}$ choices.}
\label{fig::pMtest}
\end{figure}

Finally, we want to stress that the uncertainty in $p_\mathrm{max}$ determined from real observational data is not likely to have significant effect on the derived inclination angle $\gamma$. Using the case of Vela C as an example: in Section~\ref{sec::velac}, we presented the PDF of polarization fraction (in linear space) of Vela C in Figure~\ref{fig::velacmap} (top panel). We picked $p_\mathrm{max} = 0.15$ based on the PDF, and argued that the derived $\gamma_\mathrm{obs}$ is relatively insensitive to the value of $p_\mathrm{max}$. This is demonstrated in Figure~\ref{fig::pMtest}, where we compare the results in Figure~\ref{fig::velacrobs} ($p_\mathrm{max} = 0.15$; left panel) with a different but also reasonable choice of $p_\mathrm{max}$, $0.2$ (right panel). One can clearly see that the resulting $\gamma_\mathrm{obs}$ is almost unaffected by the slight difference of $p_\mathrm{max}$. This confirms our statement above that even in the situation that $p_\mathrm{max}$ cannot be accurately recovered, our method still returns good estimates (within $\sim 30^\circ$) of the inclination angle of cloud-scale magnetic field using observed polarimetric information. 

We caution the readers again that even though our method suggests a way to estimate $p_0$ from the observed $p_{\rm max}$ that can in principle be used to derive dust grain properties at different physical scales, under some rare circumstances (when the magnetic field in the cloud is relatively well-ordered with the mean field closer to the observer's line of sight) the observed $p_{\rm max}$ may be significantly below the intrinsic maximum polarization fraction. Only in this regime the derived value of $p_0$ based on the observed $p_{\rm max}$ may not be considered as the real polarization coefficient.

\begin{table*}
	\centering
	\caption{Comparing results from models with various turbulence levels and different cloud-scale magnetic field direction. The polarization coefficient adopted in these synthetic observations is $p_0 = 0.1$. }
	\label{rotL1L10L20}
	\begin{tabular}{l | ccc | ccc | ccc | ccc } 
		\hline
		\multirow{3}{*}{Model} & \multicolumn{3}{c|}{$\gamma_{\overline{\mathbf{B}}} \approx 0^\circ$} & \multicolumn{3}{c|}{$\gamma_{\overline{\mathbf{B}}} \approx 30^\circ$} & \multicolumn{3}{c|}{$\gamma_{\overline{\mathbf{B}}} \approx 60^\circ$} & \multicolumn{3}{c}{$\gamma_{\overline{\mathbf{B}}} \approx 90^\circ$} \\
		\cline{2-13}
		 & derived & \multirow{2}{*}{${\gamma_\mathrm{3D}}^\wedge$} & \multirow{2}{*}{${\gamma_{\mathrm{obs}_,\ {\cal S}<\langle{\cal S}\rangle}}^\wedge$} & derived & \multirow{2}{*}{${\gamma_\mathrm{3D}}^\wedge$} & \multirow{2}{*}{${\gamma_{\mathrm{obs}_,\ {\cal S}<\langle{\cal S}\rangle}}^\wedge$} & derived & \multirow{2}{*}{${\gamma_\mathrm{3D}}^\wedge$} & \multirow{2}{*}{${\gamma_{\mathrm{obs}_,\ {\cal S}<\langle{\cal S}\rangle}}^\wedge$} & derived & \multirow{2}{*}{${\gamma_\mathrm{3D}}^\wedge$} & \multirow{2}{*}{${\gamma_{\mathrm{obs}_,\ {\cal S}<\langle{\cal S}\rangle}}^\wedge$} \\
		 & $p_0$ & & & $p_0$ & & & $p_0$ & & & $p_0$ & & \\
		\hline
		L1 &  0.098 & 1.0 & 19.5 & 0.096 & 27.6 & 33.5 & 0.082 & 55.6 & 56.4 & 0.061 & 78.8 & 72.9 \\
		L10 & 0.095 & 1.3 & 29.6 & 0.096 & 33.1 & 41.4 & 0.096 & 55.6 & 53.5 & 0.093 & 67.2 & 67.5 \\
		L20 & 0.098 & 1.8 & 37.7 & 0.096 & 24.0 & 45.4 & 0.096 & 48.2 & 50.5 & 0.093 & 60.9 & 57.8 \\
		\hline
	\end{tabular}
\end{table*}

\bsp	
\label{lastpage}
\end{document}